\documentclass[11pt,twoside]{jcap}
\usepackage{epsfig}
\usepackage{amssymb}
\usepackage{fancyheda}
\usepackage{times}

\expandafter\let\csname equation*\endcsname\relax
\expandafter\let\csname endequation*\endcsname\relax
\usepackage{amsmath}
\usepackage{latexsym}
\usepackage{slashed,cite}
\usepackage{upgreek}
\numberwithin{equation}{section}
\usepackage{rotating}
\usepackage{amsfonts}
\usepackage{amsbsy}
\usepackage{amscd}
\usepackage{bbm}
\usepackage{titlesec}
\titlelabel{\thetitle\quad}

\newcommand{\crefs}[1]{Refs.~\cite{#1}}

\newcommand{\arcsinh}{\ensuremath{{\rm arcsinh}}}

\newcommand{\bal}{\begin{align}}
\newcommand{\eal}{\end{align}}
\newcommand{\beqs}{\begin{subequations}}
\newcommand{\eeqs}{\end{subequations}}
\newcommand{\eec}{\end{center}}
\newcommand{\bec}{\begin{center}}

\newcommand{\eem}{\end{matrix}}
\newcommand{\bem}{\begin{matrix}}
\newcommand{\eeq}{\end{equation}}
\newcommand{\beq}{\begin{equation}}
\newcommand{\ba}{\begin{array}}
\newcommand{\ea}{\end{array}}
\newcommand{\bea}{\begin{eqnarray}}
\newcommand{\eea}{\end{eqnarray}}
\newcommand{\baq}{\begin{eqnarray}}
\newcommand{\eaq}{\end{eqnarray}}

\newcommand\eqs[2]{Eqs.~(\ref{#1}) and (\ref{#2})}

\newcommand\eqss[3]{Eqs.~(\ref{#1}), (\ref{#2}) and (\ref{#3})}

\newcommand{\ftn}{\footnotesize}
\newcommand{\nsz}{\normalsize}
\newcommand{\ssz}{\scriptsize}

\newcommand{\GeV}{{\mbox{\rm GeV}}}

\newcommand{\sFref}[2]{Fig.~\ref{#1}-{\ftn\sf ({#2})}}
\newcommand{\sEref}[2]{Eq.~(\ref{#1}{\ftn\sf {#2}})}

\newcommand{\Vhi}{\ensuremath{\widehat V_{\rm I}}}

\newcommand{\dV}{\ensuremath{\Delta\widehat V_{\rm I}}}

\def\to{\rightarrow}

\def\lf{\left(}
\def\rg{\right)}

\newcommand{\mP}{\ensuremath{m_{\rm P}}}

\newcommand{\ld}{\ensuremath{\lambda}}

\newcommand{\kp}{\ensuremath{\kappa}}

\newcommand\vev[1]{\langle {#1} \rangle}

\newcommand{\Vjhi}{\ensuremath{V_{\rm I}}}
\newcommand{\Hhi}{\ensuremath{\widehat H_{\rm I}}}
\newcommand{\Khi}{\ensuremath{K}}

\newcommand{\Omg}{\ensuremath{\Omega}}
\newcommand{\Ohi}{\ensuremath{\Omega}}

\newcommand{\ns}{\ensuremath{n_{\rm s}}}
\newcommand{\as}{\ensuremath{\alpha_{\rm s}}}
\newcommand{\As}{\ensuremath{A_{\rm s}}}
\newcommand{\Ns}{\ensuremath{\widehat N_{\star}}}
\newcommand{\na}{\ensuremath{{n_{11}}}}
\newcommand{\nb}{\ensuremath{n_{2}}}
\newcommand{\nc}{\ensuremath{n_{3}}}

\newcommand{\ca}{\ensuremath{c_\mathcal{R}}}
\newcommand{\ck}{\ensuremath{c_T}}

\newcommand{\fr}{\ensuremath{f_\mathcal{R}}}
\newcommand{\hi}{\ensuremath{h_i}}

\newcommand{\fk}{\ensuremath{f_{\rm K}}}
\newcommand{\fp}{\ensuremath{f_\phi}}
\newcommand{\ft}{\ensuremath{f_\phi}}

\newcommand{\ks}{\ensuremath{k_\star}}

\newcommand{\kx}{\ensuremath{k_S}}

\newcommand{\rcc}{\ensuremath{\mathcal{R}}}
\newcommand{\rce}{\ensuremath{\widehat{\mathcal{R}}}}
\newcommand{\Ve}{\ensuremath{\widehat{V}}}
\newcommand{\He}{\ensuremath{\widehat{H}}}

\newcommand{\msn}{\ensuremath{\what m_{\rm \dph}}}

\newcommand{\Trh}{\ensuremath{T_{\rm rh}}}
\newcommand{\sg}{\ensuremath{\phi}}

\newcommand{\sgx}{\ensuremath{\phi_\star}}

\newcommand{\sgf}{\ensuremath{\phi_{\rm f}}}

\newcommand{\Ld}{\ensuremath{\Lambda_{\rm UV}}}
\newcommand{\se}{\ensuremath{\widehat\phi}}

\newcommand{\sex}{\ensuremath{\widehat{\phi}_\star}}
\newcommand{\sef}{\ensuremath{\widehat{\phi}_{\rm f}}}
\newcommand{\geu}{\ensuremath{\widehat g}}
\newcommand{\eph}{\ensuremath{\widehat \epsilon}}
\newcommand{\ith}{\ensuremath{\widehat \eta}}

\def\aal{{\bar\alpha}}
\def\bbet{{\bar\beta}}
\def\al{{\alpha}}
\def\bt{{\beta}}

\def\K{{\widehat{K}}}

\def\th{{\theta}}

\newcommand{\kca}{\ensuremath{K_{3i}}}
\newcommand{\kba}{\ensuremath{K_{2i}}}

\newcommand{\diag}{\mbox{\sf\ftn diag}}

\newcommand{\dphi}{\ensuremath{\what{\delta\phi}}}
\newcommand{\hphi}{\ensuremath{\what{\phi}}}
\newcommand{\dph}{\ensuremath{\delta\phi}}
\newcommand{\what}{\ensuremath{\widehat}}
\newcommand{\wtilde}{\ensuremath{\widetilde}}

\newcommand{\Qef}{\ensuremath{\Lambda_{\rm UV}}}

\def\trns{transplanckian}
\def\Ka{K\"{a}hler potential}
\def\Km{K\"{a}hler manifold}
\def\Kaa{K\"{a}hler~}
\def\sub{subplanckian}

\def\str{Starobinsky}
\def\FHI{IGI~}
\def\bcp{{\sc\small Bicep2}/{\it Keck Array}}
\newcommand{\plk}{{\it Planck}}

\renewenvironment{subequations}{%
\refstepcounter{equation}%
\setcounter{parentequation}{\value{equation}}%
  \setcounter{equation}{0}
  \def\theequation{\thesection.\theparentequation{\sf\ftn\alph{equation}}}%
  \ignorespaces
}{%
  \setcounter{equation}{\value{parentequation}}%
  \ignorespacesafterend
}

\begin{document}


\title{\LARGE \bfseries\scshape  Starobinsky Inflation: From non-SUSY to SUGRA Realizations}

\author{\large \bfseries\scshape Constantinos Pallis \& Nicolaos Toumbas}

\address{Department of Physics, University of Cyprus, \\ P.O. Box 20537,
Nicosia 1678, CYPRUS}

\eads{cpallis@ucy.ac.cy, nick@ucy.ac.cy}

\begin{abstract}{\vspace{5pt}\par

We review the realization of \str-type inflation within
induced-gravity \emph{Supersymmetric} ({\ftn \sf SUSY}) and
non-SUSY models. In both cases, inflation is in agreement with the
current data and can be attained for subplanckian values of the
inflaton. The corresponding effective theories retain perturbative
unitarity up to the Planck scale and the inflaton mass is
predicted to be $3\cdot10^{13}~\GeV$. The supergravity embedding
of these models is achieved by employing two gauge singlet chiral
supefields, a superpotential that is uniquely determined by a
continuous R and a discrete $\mathbb{Z}_n$ symmetry, and several
(semi)logarithmic \Ka s that respect these symmetries. Checking
various functional forms for the non-inflaton accompanying field
in the \Ka s, we identify four cases which stabilize it without
invoking higher order terms. }

\end{abstract}

\keyw{Cosmology, Modified Gravity, Supersymmetric models,
Supergravity} 


\maketitle

\tableofcontents \vskip0.10cm\noindent\rule\textwidth{.4pt}

\setcounter{page}{1} \pagestyle{fancyplain}


\rhead[\fancyplain{}{ \bf \thepage}]{\fancyplain{}{\sl Starobinsky
Inflation: From non-SUSY to SUGRA Realizations}}
\lhead[{\fancyplain{}{ \sl \leftmark}}]{\fancyplain{}{\bf
\thepage}} \cfoot{}

\section{Introduction}\label{intro}

The idea that the universe underwent a period of exponential
expansion, called inflation \cite{guth}, has proven useful not
only for solving the horizon and flatness problems of standard
cosmology, but also for providing an explanation for the scale
invariant perturbations, which are responsible for generating the
observed anisotropies in the \emph{Cosmic Microwave Background}
({\ftn\sf CMB}). One of the first incarnations of inflation is due
to Starobinsky. To date, this attractive scenario remains
predictive, since it passes successfully all the observational
tests \cite{gws,plin}. Starobinsky considered adding an $\rcc^2$
term, where $\rcc$ is the Ricci scalar, to the standard Einstein
action in order to source inflation. Recall that gravity theories
based on higher powers of $\rcc$ are equivalent to standard
gravity theories with one additional scalar degree of freedom --
see e.g. \cite{nick1}. As a result, Starobinsky inflation is
equivalent to inflation driven by a scalar field with a suitable
potential and so, it admits several interesting realizations
\cite{ketov,ketov1,R2r,eno5,eno7,zavalos,tamvakis,linde,matterlike,ellis}.

Following this route, we show in this work that
\emph{induced-gravity inflation} ({\sf\ftn IGI})
\cite{gian,nIG,rena,old,higgsflaton} is effectively \str-like,
reproducing the structure and the predictions of the original
model. Within IGI, the inflaton exhibits a strong coupling to
$\rcc$ and the reduced Planck scale is dynamically generated
through the \emph{vacuum expectation value} ({\ftn\sf v.e.v.}) of
the inflaton at the end of inflation. Therefore, the inflaton
acquires a higgs-like behavior as in theories of induced gravity
\cite{zee, higgsflaton,jones}. Apart from being compatible with
data, the resulting theory respects perturbative unitarity up to
the Planck scale \cite{gian,R2r,nIG}. Therefore, no concerns about
the validity of the corresponding effective theory arise. This is
to be contrasted with models of \emph{non-minimal inflation}
({\sf\small nMI}) \cite{sm1,nmi, linde1, atroest,nmH,quad} based
on a $\sg^n$ potential with negligible v.e.v. for the inflaton
$\sg$. Although these models yield similar observational
predictions with the \str\ model, they admit an \emph{ultraviolet}
({\ftn\sf UV}) scale well below $\mP$ for $n>2$, leading to
complications with naturalness \cite{riotto,cutoff}.

Nonetheless, IGI allows us to embed Starobinsky inflation within
${\cal N}=1$ \emph{Supergravity} ({\ftn\sf SUGRA}) in an elegant
way. The embedding is achieved by incorporating two chiral
superfields, a modulus-like field $T$ and a matter-like field $S$
appearing in the superpotential, $W$, as well as various \Ka s,
$K$, consistent with an $R$ and discrete $\mathbb{Z}_n$ symmetries
\cite{R2r, nIG, su11} -- see also \crefs{linde, eno7, zavalos,
tamvakis, rena}. In some cases \cite{eno7, R2r, nIG, su11}, the
employed $K$'s parameterize specific \Km s,  which appear in
no-scale models \cite{noscale,lahanas}. Moreover, this scheme
ensures naturally a low enough reheating temperature, potentially
consistent with the gravitino constraint
\cite{R2r,rehEllis,rehyoko} if connected with a version of the
\emph{Minimal SUSY Standard Model} ({\ftn\sf MSSM}).


An important issue in embedding IGI in SUGRA is the stabilization
of the matter-like field $S$. Indeed, when $K$ parameterizes the
$SU(2,1)/(SU(2)\times U(1))$ \Km\ \cite{eno7,linde}, the
inflationary trajectory turns out to be unstable \emph{with
respect to} ({\ftn\sf w.r.t.}) the fluctuations of $S$. This
difficulty can be overcome by adding a sufficiently large term
$\kx|S|^4$, with $\kx>0$ and $|\kx|\sim1$, in the logarithmic
function appearing in $K$, as suggested in \cref{lee} for models
of non-minimal (chaotic) inflation \cite{linde1} and applied in
\crefs{nmH,quad,nMkin,nmHkin}. This solution, however, deforms
slightly the \Km\ \cite{nick}. More importantly, it violates the
predictability of \str\ inflation, since mixed terms
$k_{ST}|S|^2|T|^2$ with $k_{ST}\geq0.01$, which can not be ignored
(without tuning), have an estimable impact \cite{talk,nIG,np1} on
the dynamics and the observables. Moreover, this solution becomes
complicated when more than two fields are considered, since all
quartic terms allowed by symmetries have to be considered, and the
analysis of the stabilization mechanism becomes tedious -- see
e.g. \crefs{nIG,np1,talk}. Alternatively, it was suggested to use
a nilpotent superfield $S$ \cite{nil}, or a charged field under a
gauged R symmetry \cite{nick}.

In this review, we revisit the issue of stabilizing $S$,
disallowing terms of the form $|S|^{2m}$, $m>1$, without caring
much about the structure of the \Km. Namely, we investigate
systematically several functions $\hi(|S|^2)$ (with $i=1,...,11$)
that appear in the choices for $K$, and we find four acceptable
forms that lead to the stabilization of $S$ during and after IGI.
The output of this analysis is new, providing results that did not
appear in the literature before. More specifically, we consider
two principal classes of $K$'s, $K_{3i}$ and $K_{2i}$,
distinguished by whether $\hi$ and $T$ appear in the same
logarithmic function. The resulting inflationary scenaria are
almost indistinguishable. The case considered in \cref{su11} is
included as one of the viable choices in the $K_{2i}$ class.
Contrary to \cref{su11}, though, we impose here the same
$\mathbb{Z}_n$ symmetry on $W$ and $K$. Consequently, the relevant
expressions for the mass spectrum and the inflationary observables
get simplified considerably compared to those displayed in
\cref{su11}. As in the non-SUSY case, IGI may be realized using
\sub\ values for the initial (non-canonically normalized) inflaton
field. The radiative corrections remain under control and
perturbative unitarity is not violated up to $\mP$ \cite{R2, nIG,
su11}, consistently with the consideration of SUGRA as an
effective theory.

Throughout this review we focus on the standard $\Lambda$CDM
cosmological model \cite{plin}. An alternative framework is
provided by the running vacuum models \cite{rvm} which turn out to
yield a quality fit to observations, significantly better than
that of $\Lambda$CDM. In this case, the acceleration of the
universe, either during inflation or at late times, is not
attributed to a scalar field but rather arises from the
modification of the vacuum itself, which is dynamical. A SUGRA
realization of Starobinsky inflation within this setting is
obtained in the last paper of \cref{ketov1}.

The plan of this paper is as follows. In Sec.~\ref{sec:sti}, we
establish the realization of Starobinsky inflation as IGI in a
non-SUSY framework. In \Sref{sec:sugra} we introduce the
formulation of IGI in SUGRA and revisit the issue of stabilizing
the matter-like field $S$. The emerging inflationary models are
analyzed in Sec.~\ref{sec:inf}. Our conclusions are summarized in
Sec.~\ref{sec:con}. Throughout, charge conjugation is denoted by a
star ($^*$), the symbol $,z$ as subscript denotes derivation
w.r.t. $z$, and we use units where the reduced Planck scale, $\mP
= 2.43\cdot 10^{18}~\GeV$, is set equal to unity.

\section{Starobinsky Inflation From Induced Gravity}\label{sec:sti}

We begin our presentation demonstrating the connection between
$\rcc^2$ inflation and IGI. We first review the formulation of nMI
in \Sref{fhi}, and then proceed to describe the inflationary
analysis in \Sref{obs}. Armed with these prerequisites, we present
$\rcc^2$ inflation as a type of nMI in \Sref{r2nmi}, and exhibit
its connection with IGI in \Sref{igi}.

\subsection{Coupling non-Minimally the Inflaton to Gravity}\label{fhi}

We consider an inflaton $\sg$ that is non-minimally coupled to the
Ricci scalar $\rcc$, via a coupling function $\fr(\sg)$. We denote
the inflaton potential by $V_{\rm I}(\sg)$ and allow for a general
kinetic function $\fk(\sg)$ -- in the cases of pure nMI \cite{old,
nmi, atroest} $\fk=1$. The \emph{Jordan Frame} ({\ftn\sf JF})
action takes the form
\beq \label{action1} {\sf  S} = \int d^4 x \sqrt{-\mathfrak{g}}
\left(-\frac{1}{2}\fr\rcc +\frac{1}{2}\fk g^{\mu\nu}
\partial_\mu \sg\partial_\nu \sg-
\Vjhi(\sg)\right)\,, \eeq
where $\mathfrak{g}$ is the determinant of the
Friedmann-Robertson-Walker metric, $g_{\mu\nu}$, with signature
$(+,-,-,-)$. We require $\vev{\fr}\simeq1$ to ensure
ordinary Einstein gravity at low energies.

By performing a conformal transformation \cite{nmi} to
the \emph{Einstein frame} ({\sf\ftn EF}), we write the action
\beq {\sf  S}= \int d^4 x
\sqrt{-\what{\mathfrak{g}}}\left(-\frac12
\rce+\frac12\geu^{\mu\nu} \partial_\mu \se\partial_\nu \se
-\Vhi(\se)\right), \label{action} \eeq
where a hat denotes an EF quantity. The EF metric is given by
$\geu_{\mu\nu}=\fr\,g_{\mu\nu}$, and the canonically normalized field, $\se$, and its
potential, $\Vhi$, are defined as follows:
\beq \label{VJe} \mbox{\ftn\sf (a)}\>\>\>
\frac{d\se}{d\sg}=J=\sqrt{\frac{\fk}{\fr}+{3\over2}\left({{\fr}_{,\sg}\over \fr}\right)^2}\>\>\>\mbox{and}\>\>\>\mbox{\ftn\sf
(b)}\>\>\> \Vhi= \frac{\Vjhi}{\fr^2}\,.\eeq
For $\fr\gg\fk$, the coupling function $\fr$ acquires a twofold
role. On one hand, it determines the relation between $\se$ and
$\sg$. On the other hand, it controls the shape of $\Vhi$, thus
affecting the observational predictions -- see below. The analysis
of nMI can be performed in the EF, using the standard slow-roll
approximation. It is \cite{old} completely equivalent with the
analysis in the JF. We just have to keep track the relation
between $\se$ and $\sg$.

\subsection{Observational and Theoretical Constraints} \label{obs}

A viable model of nMI must be compatible with a number of
observational and theoretical requirements summarized in the
following -- cf. \cref{review}.

\paragraph{\bf 1.2.1} The number of e-foldings $\Ns$ that the scale $\ks=0.05/{\rm Mpc}$
experiences during inflation must to be large enough for the resolution
of the horizon and flatness problems of the standard hot Big Bang model, i.e.
\cite{nmi,plin},
\begin{equation}
\label{Nhi}  \Ns=\int_{\sef}^{\sex} d\se\frac{\Vhi}{\Ve_{\rm
I,\se}}=\int_{\sgf}^{\sgx}\, d\sg\: J^2\frac{\Vhi}{\Ve_{\rm
I,\sg}}\simeq61.7+\ln{\what V_{\rm I}(\sgx)^{1/2}\over\what V_{\rm
I}(\sgf)^{1/3}}+ {1\over3}\ln T_{\rm
rh}+{1\over2}\ln{\fr(\sgx)\over\fr(\sgf)^{1/3}}\,,\eeq
where $\sgx~[\sex]$ is the value of $\sg~[\se]$ when $\ks$
crosses the inflationary horizon. In deriving the formula above --
cf. \cref{nMkin} -- we take into account an equation-of-state with
parameter $w_{\rm rh}=0$ \cite{turner}, since $\Vhi$ can be well
approximated by a quadratic potential for low values of $\sg$ -- see
\eqss{Vestr}{Veig}{Vhiexp} below. Also $\Trh$ is the reheating
temperature after nMI. We take a representative value
$\Trh=4.1\cdot10^{-10}$ throughout, which results to $\Ns\simeq53$.
The effective number of relativistic degrees of freedom at
temperature $\Trh$ is taken $g_{\rm rh}=107.75$ in accordance with
the Standard model spectrum. Lastly, $\sg_{\rm f}~[\se_{\rm f}]$
is the value of $\sg~[\se]$ at the end of nMI, which in the slow-roll approximation
can be obtained via the condition
$$ {\ftn\sf max}\{\widehat\epsilon(\sg_{\rm
f}),|\widehat\eta(\sg_{\rm f})|\}=1,\>\>\>~\mbox{where}$$
\beq \label{sr}\widehat\epsilon= {1\over2}\left(\frac{\Ve_{\rm
I,\se}}{\Ve_{\rm I}}\right)^2={1\over2J^2}\left(\frac{\Ve_{\rm
I,\sg}}{\Ve_{\rm I}}\right)^2 \>\>\>\mbox{and}\>\>\>\widehat\eta=
\frac{\Ve_{\rm I,\se\se}}{\Ve_{\rm I}}={1\over
J^2}\left(\frac{\Ve_{\rm I,\sg\sg}}{\Ve_{\rm I}}-\frac{\Ve_{\rm
I,\sg}}{\Ve_{\rm I}}{J_{,\sg}\over J}\right)\cdot \eeq
Evidently non trivial
modifications of $\fr$, and thus of $J$, may have a significant effect
on the parameters above, modifying the inflationary
observables.

\paragraph{\bf 1.2.2} The amplitude $\As$ of the power spectrum of the curvature perturbation
generated by $\sg$ at $k_{\star}$ has to be consistent with the
data~\cite{plcp}, i.e.,
\begin{equation}  \label{Prob}
\sqrt{\As}=\: \frac{1}{2\sqrt{3}\, \pi} \; \frac{\Ve_{\rm
I}(\sex)^{3/2}}{|\Ve_{\rm I,\se}(\sex)|}
=\frac{|J(\sgx)|}{2\sqrt{3}\, \pi} \; \frac{\Ve_{\rm
I}(\sgx)^{3/2}}{|\Ve_{\rm I,\sg}(\sgx)|}\simeq4.627\cdot
10^{-5}\,.
\end{equation}
As shown in \Sref{stab}, the remaining scalars in the SUGRA
versions of nMI may be rendered heavy enough and so, they do not
contribute to $\As$.

\paragraph{\bf 1.2.3} The remaining inflationary observables (the spectral index $\ns$,
its running $\as$, and the tensor-to-scalar ratio $r$) must be in
agreement with the fitting of the \plk, \emph{Baryon Acoustic
Oscillations} ({\sf\ftn BAO}) and \bcp\ data \cite{plin,gws} with the
$\Lambda$CDM$+r$ model, i.e.,
\begin{equation}  \label{nswmap}
\mbox{\ftn\sf
(a)}\>\>\ns=0.968\pm0.009\>\>\>\mbox{and}\>\>\>\mbox{\ftn\sf
(b)}\>\>r\leq0.07,
\end{equation}
at the 95$\%$ \emph{confidence level} ({\sf\ftn c.l.}) with
$|\as|\ll0.01$. Although compatible with \sEref{nswmap}{b}, all
data taken by the \bcp\ CMB polarization experiments, up to the
2014 observational season ({\sf\ftn BK14}) \cite{gws}, seem to
favor $r$'s of the order of $0.01$, as the reported value is
$0.028^{+0.026}_{-0.025}$ at the 68$\%$ c.l.. These inflationary
observables are estimated through the relations:
\beq\label{ns} \mbox{\ftn\sf (a)}\>\>\> \ns=\: 1-6\eph_\star\ +\
2\ith_\star,\>\>\>\mbox{\ftn\sf (b)}\>\>\>
\as=\:\frac23\left(4\widehat\eta_\star^2-(\ns-1)^2\right)-2\widehat\xi_\star\>\>\>\mbox{and}\>\>\>\mbox{\ftn\sf
(c)}\>\>\>r=16\eph_\star, \eeq
where $\widehat\xi={\Ve_{\rm I,\se} \Ve_{\rm I,\se\se\se}/\Ve_{\rm
I}^2}$ and the variables with subscript $\star$ are evaluated at
$\sgx$.

\paragraph{\bf 1.2.4} The effective theory describing nMI remains
valid up to a UV cutoff scale $\Qef$, which has to be large enough to ensure the stability of
our inflationary solutions, i.e.,
\beq \label{subP}\mbox{\ftn\sf (a)}\>\> \Vhi(\sgx)^{1/4}\leq\Qef
\>\>\>~\mbox{and}\>\>\>~\mbox{\ftn\sf (b)}\>\>\sgx\leq\Qef.\eeq
As we show below, $\Qef\simeq1$ for the models analyzed in this
work, contrary to the cases of pure nMI with large $\fr$, where
$\Qef\ll1$. The determination of $\Qef$ is achieved expanding
${\sf S}$ in \Eref{action} about $\vev{\phi}$. Although these
expansions are not strictly valid \cite{riotto} during inflation,
we take the $\Qef$ extracted this way to be the overall UV cut-off
scale, since the reheating phase -- realized via oscillations
about $\vev{\sg}$ -- is a necessary stage of the inflationary
dynamics.

\subsection{From Non-Minimal to $\rcc^2$ Inflation}\label{r2nmi}

The $\rcc^2$ inflation can be viewed as a type of nMI, if we
employ an auxiliary field $\sg$ with the following input
ingredients
\beq \fk=0,~~\fr=1+4\ca \sg~~~\mbox{and}~~~\Vhi=\sg^2\,.
\label{str1}\eeq
Using the equation of motion for the auxiliary field, $\sg=\ca
\rcc$, we obtain the action of the original Starobinsky model (see
e.g. \cref{nick}):
\beq \label{strS} {\sf S} = \int d^4 x \sqrt{-\mathfrak{g}}
\left(-\frac12\rcc+\ca ^2\rcc^2\right). \eeq
As we can see from \Eref{str1}, the model has only one free parameter
($\ca$), enough to render it consistent with the observational data, ensuring at the same time
perturbative unitarity up to the Planck scale. Using
\Eref{str1} and \Eref{VJe}, we obtain the EF quantities
\beq \mbox{\ftn\sf (a)}\>\>
J=2\sqrt{6}\frac{\ca}{\fr}~~~\mbox{and}~~~\mbox{\ftn\sf (b)}\>\>
\Vhi={\sg^2\over\fr^2}\simeq{1\over16\ca^2}\,\cdot\label{VJstr}\eeq
For $\ca\gg1$, the plot of $\Vhi$ versus $\sg$ is depicted in
\sFref{fig1}{a}. An inflationary era can be supported since $\Vhi$
becomes flat enough. To examine further this possibility, we
calculate the slow-roll parameters. Plugging \Eref{VJstr} into
\Eref{sr} yields
\beq \label{srstr}\widehat\epsilon={1\over12\ca
^2\sg^2}\>\>\>\mbox{and}\>\>\> \widehat\eta={1-4\ca\sg\over12\ca^2
\sg^2}\cdot \eeq
Notice that $\eta<0$ since $\Vhi$ is slightly concave downwards, as shown
in \sFref{fig1}{a}. The value of $\sg$ at the end of nMI is determined via \Eref{sr}, giving
\beq \label{sgfstr} \sgf=\mbox{\ftn\sf
max}\lf{1\over2\sqrt{3}\ca},{1\over6\ca}\rg\>\Rightarrow\>
\sgf={1\over2\sqrt{3}\ca}\cdot\eeq

Under the assumption that $\sgf\ll\sgx$, we can obtain a relation between $\Ns$ and $\sgx$ via
\Eref{Nhi}
\beq\label{sgxstr} \Ns\simeq{3\ca \sgx}.\eeq The precise value of
$\ca$ can be  determined enforcing \Eref{Prob}. Recalling that
$\Ns\simeq53$, we get
\beq \label{Probstr} \As^{1/2}\simeq {\Ns\over12\sqrt{2}\pi\ca
}=4.627\cdot10^{-5} ~~\Rightarrow~~\ca\simeq2.3\cdot10^4.\eeq 
The resulting value of $\ca$ is large enough so that
\beq \sgx\simeq{\Ns/3\ca}\simeq8.3\cdot10^{-4}\ll1 \eeq
consistently with \sEref{subP}{b} -- see \sFref{fig1}{a}.
Impressively, the remaining observables turn out to be compatible
with the observational data of \Eref{nswmap}. Indeed, inserting
the above value of $\sgx$ into \Eref{ns} ($\Ns=53$), we get
\beqs\bea
\label{nsstr}\ns&\simeq&\frac{(\Ns-3)(\Ns-1)}{\Ns^2}\simeq1-\frac{2}{\Ns}-\frac{3}{\Ns^2}\simeq0.961;\\
\label{asstr} \as&\simeq&-\frac{(\Ns-3)(4\Ns+3)}{2\Ns^4}
\simeq-{2\over\Ns^2}-{15\over2\Ns^3}\simeq-7.6\cdot10^{-4};\\
\label{rstr} r&\simeq&{12\over\Ns^2}\simeq4.2\cdot
10^{-3}\,.\eea\eeqs
Without the simplification of \Eref{sgxstr}, we obtain
numerically $\ns=0.964$, $\as=-6.7\cdot10^{-4}$ and
$r=3.7\cdot10^{-3}$. We see that $\ns$ turns out to be appreciably
lower than unity thanks to the negative values of $\eta$ -- see \Eref{srstr}.
The mass of the inflaton at the vacuum is
\beq \msn=\left\langle\Ve_{\rm I,\se\se}\right\rangle^{1/2}=
\left\langle \Ve_{\rm
I,\sg\sg}/J^2\right\rangle^{1/2}={1/2\sqrt{3}\ca
}\simeq1.25\cdot10^{-5}~~(\mbox{i.e.}~~3\cdot10^{13}~\GeV).\label{msstr}\eeq
As we show below this value is a salient future in all models
of Starobinsky inflation.

Furthermore, the model provides an elegant solution to the
unitarity problem \cite{riotto,cutoff}, which plagues models of
nMI with $\fr\sim\sg^n\gg\fk$, $n>2$ and $\fk=1$. This stems from
the fact that $\hphi$ and $\sg$ do not coincide at the vacuum, as
\sEref{VJstr}{a} implies $\hphi=\vev{J}\sg=2\sqrt{3}\ca\phi$. In
fact, if we expand the second term in the \emph{right-hand side}
({\sf\ftn r.h.s.}) of \Eref{action} about $\vev{\sg}=0$, we find
\beqs\beq \label{Jstr} J^2
\dot\phi^2=\lf1-2\sqrt{\frac{2}{3}}{\hphi}+2{\hphi^2}-\cdots\rg\dot{\hphi}^2\,.\eeq
Similarly, expanding $\Vhi$ in \sEref{VJstr}{b}, we obtain
\beq \label{Vestr}
\Vhi=\frac{\hphi^2}{24\ca^2}\lf1-2\sqrt{\frac{2}{3}}{\hphi}+
2{\hphi^2}-\cdots\rg\,.\eeq\eeqs
Since the coefficients of the above series are of order unity,
independent of $\ca$, we infer that the model does not face any
problem with perturbative unitarity up to the Planck scale.

\begin{figure}[!t]\vspace*{-.09in}
\hspace*{-.15in}
\begin{minipage}{8in}
\epsfig{file=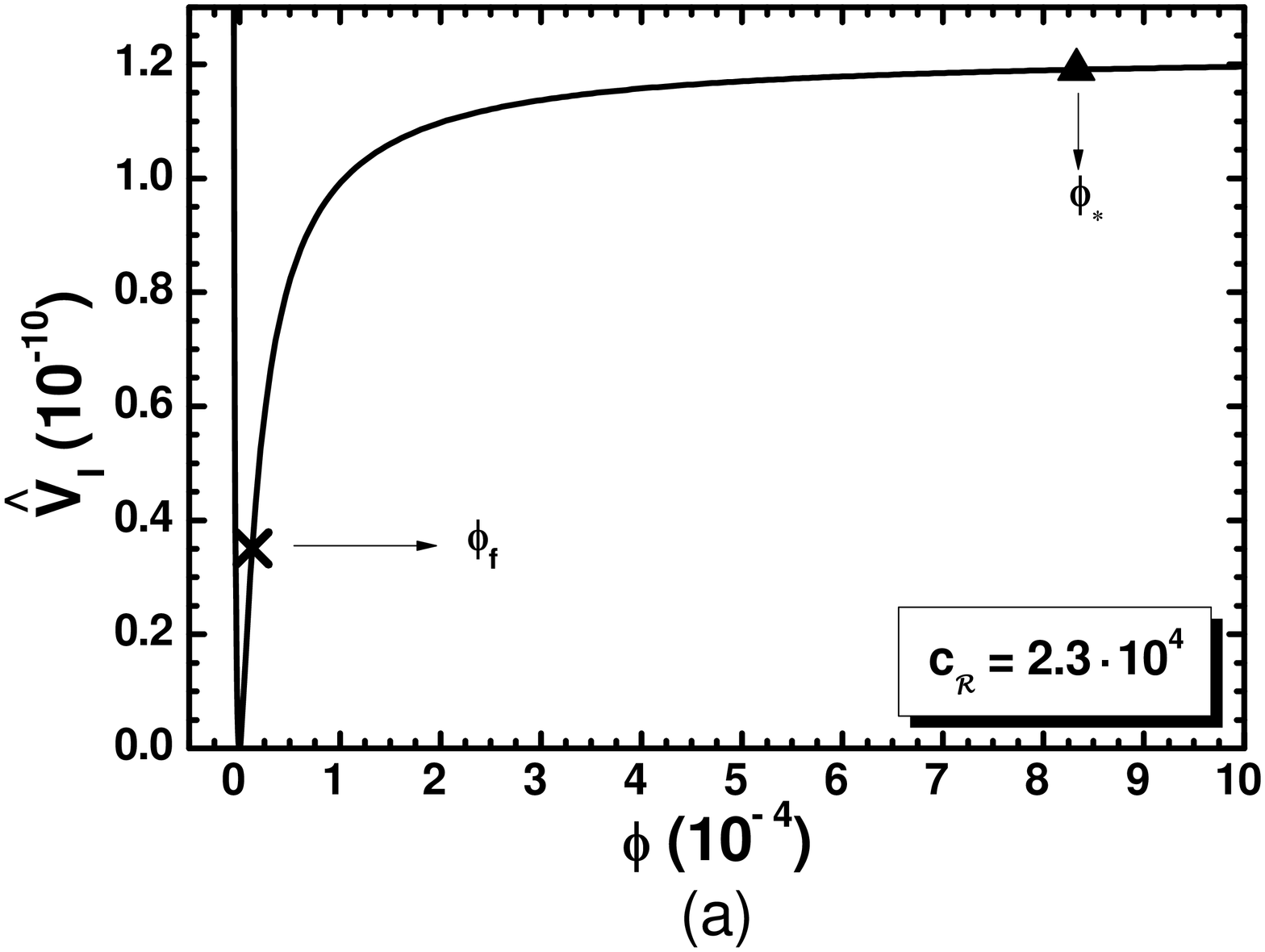,height=3.60in,angle=-90}
\hspace*{-1.37 cm}
\epsfig{file=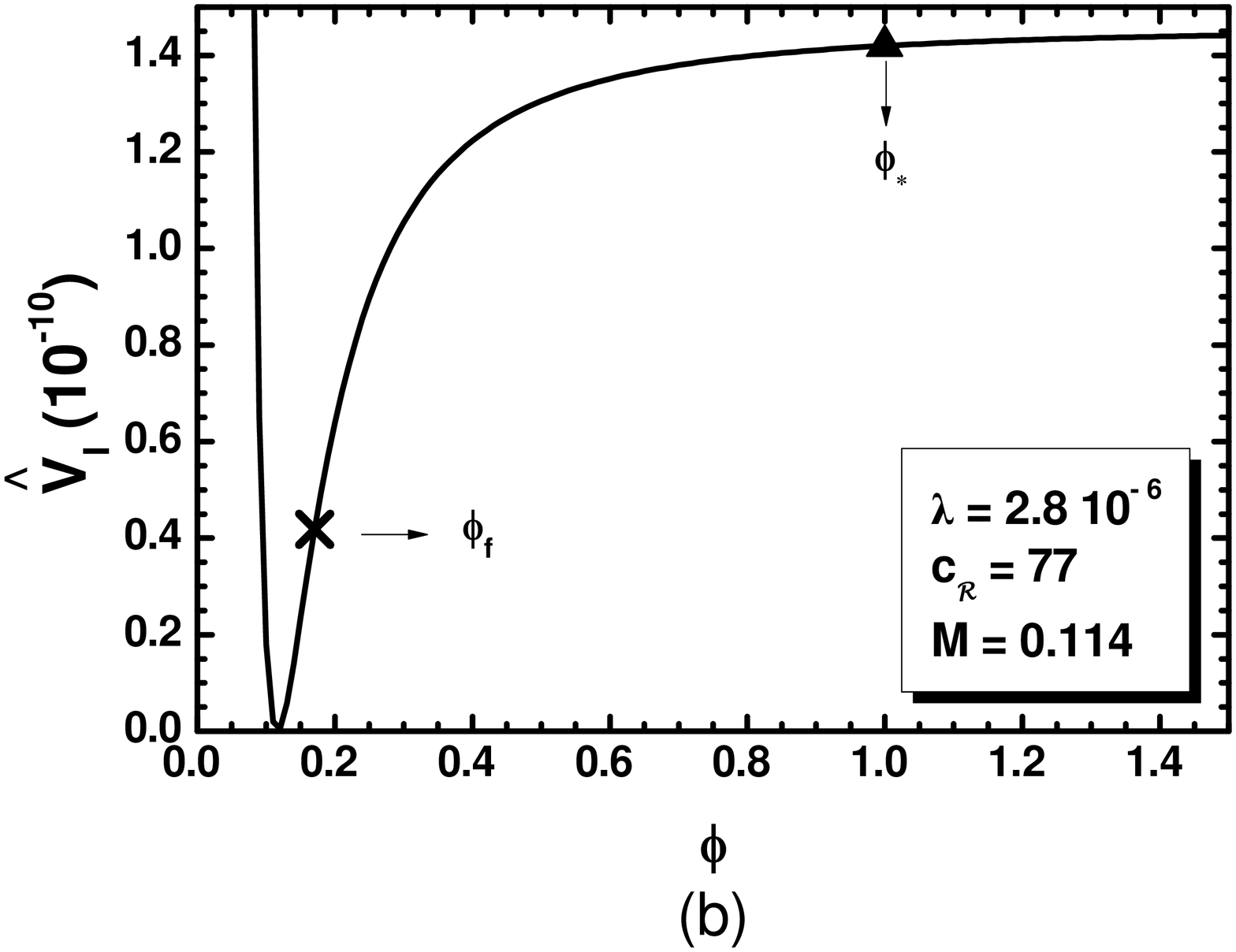,height=3.60in,angle=-90} \hfill
\end{minipage}
\hfill \caption[]{\sl The inflationary potential $\Vhi$ as a
function of $\sg$ for $\rcc^2$ inflation (${\sf\ftn a}$) and IGI
with $\sgx=1$ (${\sf\ftn b}$). Values corresponding to $\sgx$ and
$\sgf$ are also indicated.}\label{fig1}
\end{figure}

\subsection{Induced-Gravity Inflation}\label{igi}

It would be certainly beneficial to realize the structure and the
predictions of $\rcc^2$ inflation in a framework that deviates
minimally from Einstein gravity, at least in the present
cosmological era. To this extent, we incorporate the idea of
induced gravity, according to which $\mP$ is generated dynamically
\cite{zee} via the v.e.v. of a scalar field $\sg$, driving a phase
transition in the early universe. The simplest way to implement
this scheme is to employ a double-well potential for $\sg$ -- for
scale invariant realizations of this idea see \cref{jones}. On the
other hand, an inflationary stage requires a sufficiently flat
potential, as in \Eref{str1}. This can be achieved at large field
values if we introduce a quadratic $\fr$ \cite{old,higgsflaton}.
More explicitly, IGI may be defined as a nMI with the following
input ingredients:
\beq \fk=1,~~\fr=\ca\sg^2~~\mbox{and}~~V_{\rm
I}=\ld\lf\sg^2-M^2\rg^2/4\,.\label{ig1}\eeq
Given that  $\vev{\sg}=M$, we recover Einstein gravity at the
vacuum if \beq
\fr(\vev{\sg})=1~~\Rightarrow~~M=1/\sqrt{\ca}.\label{ig2} \eeq
We see that in this model there is one additional free parameter, namely $\ld$ appearing in the potential, as compared
to the $\rcc^2$ model.

\Eref{VJe} and \Eref{ig1} imply
\beq \mbox{\ftn\sf (a)}\>\>J\simeq{\sqrt{6}/\sg}
~~\mbox{and}~~\mbox{\ftn\sf (b)}\>\>\Vhi=\frac{\ld\ft^2}{4 \ca^4
\sg^4}\simeq \frac{\ld}{4\ca^2}~~\mbox{with}~~\ft=1-\ca
\sg^2\,.\label{VJig}\eeq For $\ca\gg1$, the plot of $\Vhi$ versus
$\sg$ is shown in \sFref{fig1}{b}. As in the $\rcc^2$ model,
$\Vhi$ develops a plateau and so, an inflationary stage can be
realized. To check its robustness, we compute the slow-roll
parameters. \Eref{sr} and \Eref{VJig} give
\beq\widehat\epsilon=\frac{4}{3
\ft^2}~~\mbox{and}~~\widehat\eta=\frac{4(1+\ft)}{3
\ft^2}\,\cdot\eeq
IGI is terminated when $\sg=\sgf$, determined by the condition
\beq \sgf=\mbox{\ftn\sf max}\lf\sqrt{1 +
2/\sqrt{3}\over\ca},\sqrt{5\over3\ca}\rg\>\Rightarrow\>
\sgf=\sqrt{1 + 2/\sqrt{3}\over\ca}\cdot\eeq

Under the assumption that $\sgf\ll\sgx$, \Eref{Nhi} implies the following relation between $\Ns$ and $\sgx$
\beq \label{sgxig} \Ns\simeq{3\ca
\sgx^2/4}\>\Rightarrow\>\sgx\simeq2\sqrt{\Ns\over3\ca}\gg\sgf.\eeq
Imposing \sEref{subP}{b} and setting $\Ns\simeq53$, we derive a lower bound on $\ca$:
\beq\sgx\leq1\>\Rightarrow\>\ca\geq4\Ns/3\simeq71.
\label{subPig}\eeq
Contrary to $\rcc^2$ inflation, $\ca$ does not control exclusively
the normalization of \Eref{Prob}, thanks to the presence of an
extra factor of $\sqrt{\ld}$. This is constrained to scale with
$\ca$. Indeed, we have
\beq \label{Probig} \As^{1/2}\simeq
{\sqrt{\ld}\Ns\over6\sqrt{2}\pi\ca }=4.627\cdot10^{-5}
\>\Rightarrow\> \ca \simeq42969\,
\sqrt{\ld}~~\mbox{for}~~\Ns\simeq53\,.\eeq If, in addition, we
impose the perturbative bound $\ld\leq3.5$, we end-up with
following ranges:
\beq \label{res4}77\lesssim\ca\lesssim8.5\cdot10^4~~\mbox{and}~~
2.8\cdot10^{-6}\lesssim\lambda\lesssim3.5\,,\eeq
where the lower bounds on $\ck$ and $\ld$ correspond to $\sgx=1$
-- see \sFref{fig1}{b}. Within the allowed ranges, $\msn$ remains
constant, by virtue of \Eref{Probig}. The mass turns out to be
\beq \msn={\sqrt{\ld}/\sqrt{3}\ca}\simeq1.25\cdot10^{-5},
\label{msig} \eeq
essentially equal to that estimated in \Eref{msstr}. Moreover,
using \Eref{sgxig} and \Eref{ns}, we extract the remaining
observables
\beqs\bea
\label{nsig}\ns&=&\frac{(4\Ns-15)(4\Ns+1)}{(3-4\Ns)^2}\simeq1-\frac{2}{\Ns}-\frac{9}{2\Ns^2}\simeq0.961;\\
\label{asig} \as&=&-\frac{128\Ns(4\Ns+9)}{(3-4\Ns)^4}
\simeq-{2\over\Ns^2}-{21\over2\Ns^3}\simeq-7.7\cdot10^{-4};\\
\label{rig}
r&=&{192\over(3-4\Ns)^2}\simeq\frac{12}{\Ns^2}\simeq4.4\cdot
10^{-3}\,.\eea\eeqs
Without making the approximation of \Eref{sgxig}, we obtain
numerically
$(\ns,\as,r)=(0.964,-6.6\cdot10^{-4},3.7\cdot10^{-3})$. These
results practically coincide with those of $\rcc^2$ inflation,
given in Eqs.~(\ref{nsstr}) -- (\ref{rstr}), and they are in
excellent agreement with the observational data presented in
\Eref{nswmap}.

As in the previous section, the model retains perturbative unitarity up to
$\mP$. To verify this, we first expand the second term in the
r.h.s. of \Eref{action1} about $\dphi=\phi-M\simeq0$, with $J$ given
by \sEref{VJig}{a}. We find
\beqs\beq \label{Jig} J^2
\dot\phi^2=\lf1-\sqrt{\frac{2}{3}}{\dphi}+\frac12{\dphi^2}-\cdots\rg\dot{\dphi}^2~~\mbox{with}~~~
\dphi\simeq\sqrt{6\ca}\dph\,.\eeq Expanding $\Vhi$
given by \sEref{VJig}{b}, we get
\beq\label{Veig}\Vhi=\frac{\ld^2}{6\ca
^2}\dphi^2\lf1-\sqrt{\frac{3}{2}}{\dphi}+
\frac{25}{24}{\dphi^2}-\cdots\rg\,.\eeq\eeqs
Therefore, $\Qef=1$ as for $\rcc^2$ inflation. Practically
identical results can be obtained if we replace the quadratic
exponents in \Eref{ig1} with $n\geq3$ as first pointed out in
\cref{gian}. This generalization can be elegantly performed
\cite{rena,nIG} within SUGRA, as we review below.

\section{Induced-Gravity Inflation in SUGRA} \label{sec:sugra}

In \Sref{sec:sugra1}, we present the general SUGRA
setting, where IGI is embedded. Then, in \Sref{sec:ks},
we examine a variety of \Ka s, which lead to the desired
inflationary potential -- see \Sref{sec:ks2}. We check the
stability of the inflationary trajectory in \Sref{stab}.

\subsection{The General Set-Up} \label{sec:sugra1}

To realize IGI within SUGRA \cite{R2r, nIG, rena,su11},
we must use of two gauge singlet chiral superfields $z^\al$,
with $z^1=T$ and $z^2=S$ being the inflaton and a ``stabilizer''
superfield respectively. Throughout this work, the complex scalar
fields $z^\al$ are denoted by the same superfield symbol. The EF
effective action is written as follows \cite{linde1}
\beqs \beq\label{Saction1}  {\sf S}=\int d^4x \sqrt{-\what{
\mathfrak{g}}}\lf-\frac{1}{2}\rce +K_{\al\bbet}
\geu^{\mu\nu}\partial_\mu z^\al \partial_\nu z^{*\bbet}-\Ve\rg\,,
\eeq
%
where $K_{\al\bbet}=K_{,z^\al z^{*\bbet}}$ is the K\"ahler metric
and $K^{\al\bbet}$ its inverse
($K^{\al\bbet}K_{\bbet\gamma}=\delta^\al_{\gamma}$). $\Ve$ is the
EF F--term SUGRA potential, given in terms of the \Ka\ $K$ and the
superpotential $W$ by the following expression
\beq \Ve=e^{\Khi}\left(K^{\al\bbet}D_\al W D^*_\bbet W^*-3{\vert
W\vert^2}\right)~~~\mbox{with}~~~D_\al W=W_{,z^\al}
+K_{,z^\al}W\,.\label{Vsugra} \eeq\eeqs

Conformally transforming to the JF
with $\fr=-{\Omega}/{N}$, where $N$ is a
dimensionless positive parameter, ${\sf S}$
takes the form
\beq {\sf S}=\int d^4x \sqrt{-\mathfrak{g}}\lf\frac{\Omega}{2N}
\rcc+\frac{3}{4N\Omega}\partial_\mu\Omega \partial^\mu\Omega
-\frac{1}{N}\Omega K_{\al{\bbet}}\partial_\mu z^\al \partial^\mu
z^{*\bbet}-V\rg\>\>\>\mbox{with}\>\>\>V=\frac{\Omg^2}{N^2}\Ve\,.\label{action2}\eeq
Note that  $N=3$ reproduces the standard set-up \cite{linde1}. Let
us also relate $\Omega$ and $K$ by
\beq-\Omega/N =e^{-K/N}\>\Rightarrow\>K=-N
\ln\lf-\Omega/N\rg\,.\label{Omg1}\eeq
Then taking into account the definition \cite{linde1} of the purely
bosonic part of the auxiliary field when on shell,
\beq {\cal A}_\mu =i\lf K_\al \partial_\mu z^\al-K_\aal
\partial_\mu z^{*\aal}\rg/6, \label{Acal1}\eeq
we arrive at the following action
\beqs\beq  {\sf S}=\int d^4x \sqrt{-\mathfrak{g}}\lf\frac{
\Omega}{2N}\rcc+\lf\Omega_{\al{\bbet}}+\frac{3-N}{N}
\frac{\Omega_{\al}\Omega_{\bbet}}{\Omega}\rg \partial_\mu z^\al
\partial^\mu z^{*\bbet}- \frac{27}{N^3}\Omega{\cal A}_\mu{\cal A}^\mu-V
\rg\,.\label{Sfinal}\eeq
By virtue of \Eref{Omg1}, ${\cal A}_\mu$ takes
the form
\beq {\cal A}_\mu =-iN \lf \Omega_\al \partial_\mu
z^\al-\Omega_\aal
\partial_\mu z^{*\aal}\rg/6\Omega\,\label{Acal}\eeq\eeqs
with $\Omega_\al=\Omega_{,z^\al}$ and
$\Omega_\aal=\Omega_{,z^{*\aal}}$. As can be seen from \Eref{Sfinal},
$-\Omega/{N}$ introduces a non-minimal coupling of the scalar fields to
gravity. Ordinary Einstein gravity is recovered at the vacuum
when
\beq -\vev{\Ohi}/N\simeq1. \label{ig}\eeq

Starting with the JF action in \Eref{Sfinal}, we seek to realize
IGI, postulating the invariance of $\Ohi$ under the action of a
global $\mathbb{Z}_n$ discrete symmetry. With $S$ stabilized at
the origin, we write
\beq -\Ohi/N=\Omega_{\rm H}(T)+\Omega^*_{\rm
H}(T^*)\>\>\>\mbox{with}\>\>\>\Omega_{\rm H}(T)= \ck
{T^n}+\sum_{k=2}^\infty\lambda_{k}{T^{kn}}, \label{Omdef}\eeq
where $k$ is a positive integer. If $T\leq1$ during IGI and assuming
that $\lambda_{k}$'s are relatively small, the contributions of the
higher powers of $T$ in the expression above are small, and
these can be dropped. As we verify later, this can be achieved
when the coefficient $c_T$ is large enough. Equivalently, we may
rescale the inflaton, setting $T \to \tilde{T}={c_T}^{1/n}T$. Then
the coefficients $\lambda_{k}$ of the higher powers in the
expression of $\Ohi$ get suppressed by factors of $c_T^{-k}$.
Thus, $\mathbb{Z}_n$ and the requirement $T\leq1$ determine the
form of $\Ohi$, avoiding a severe tuning of the coefficients
$\lambda_{k}$. Under these assumptions, $K$ in \Eref{Omg1} takes
the form
\beq K_0=-N\ln\Big(
f(T)+f^*(T^*)\Big)\>\>\>\mbox{with}\>\>\>f(T)\simeq\ck T^n \,,
\label{K0} \eeq
where $S$ is assumed to be stabilized at the origin.

\eqs{Omg1}{ig}
require that $T$ and $S$ acquire the following v.e.v.s
\beq \vev{T}\simeq1/(2\ck)^{1/n}\>\>\>\mbox{and}\>\>\>
\vev{S}=0\,.\label{vevs} \eeq
These v.e.v.s can be achieved, if we choose the following
superpotential \cite{nIG,rena}:
\beq \label{Wn} W=\ld S\lf T^n-1/2\ck\rg\,. \eeq
Indeed the corresponding F-term SUSY potential, $V_{\rm SUSY}$, is
found to be
\beq V_{\rm SUSY}= \ld^2\left|T^n- 1/2\ck\right|^2 +
\ld^2n^2\left|ST^{n-1}\right|^2 \label{VF}\eeq
and is minimized by the field configuration in \Eref{ig1}.

As emphasized in \crefs{R2r,nIG,su11}, the forms of $W$ and
$\Omega_{\rm H}$ can be uniquely determined if we limit ourselves
to integer values for $n$ (with $n>1$) and $T\leq1$, and impose two
symmetries:

\begin{itemize}
\item[\sf\ftn (i)] An R symmetry under which $S$ and $T$ have
charges $1$ and $0$ respectively;
\item[\sf\ftn (ii)] A discrete symmetry $\mathbb{Z}_n$ under which
only $T$ is charged.
\end{itemize}
For simplicity we assume here that both $W$ and $\Omega_{\rm H}$
respect the same $\mathbb{Z}_n$, contrary to the situation in
\cref{su11}. This assumption simplifies significantly the formulae
in Secs.~\ref{sec:ks2} and \ref{stab}. Note, finally, that the
selected $\Omega$ in \Eref{Omdef} does not contribute in the
kinetic term involving $\Omega_{TT^*}$ in \Eref{Sfinal}. We expect
that our findings are essentially unaltered even if we include in
the r.h.s. of \Eref{Omdef} a term $-(T-T^*)^2/2N$ \cite{rena} or
$-|T|^2/N$ \cite{nIG} which yields $\Omega_{TT^*}=1\ll\ck$ -- the
former choice, though, violates the $\mathbb{Z}_n$ symmetry above.

\subsection{Proposed K\"ahler Potentials}\label{sec:ks}

It is obvious from the considerations above, that the
stabilization of $S$ at zero during and after IGI is of crucial
importance for the viability of our scenario. This key issue can
be addressed if we specify the dependence of the \Ka\ on $S$. We
distinguish the following basic cases:
\beq\kca=-\nc\ln\Big(f(T)+f^*(T^*)+\hi(X)\Big)~~~
\mbox{and}~~~\kba=-\nb\ln\Big( f(T)+f^*(T^*)\Big)+\hi(X),
\label{k32} \eeq
where the various choices $\hi$, $i=1,...,11$, are specified in \Tref{tab1}, and $X$
is defined as follows
\beq \label{defx} X=\begin{cases}-|S|^2/\nc &\mbox{for}~~~K=\kca\\
|S|^2 &\mbox{for}~~~K=\kba\,.\end{cases}\eeq
As shown in \Tref{tab1} we consider exponential, logarithmic,
trigonometric and hyperbolic functions. Note that $K_{31}$ and
$K_{21}$ parameterize the $SU(2,1)/SU(2)\times U(1)$ and
$SU(1,1)/U(1)\times U(1)$ \Km s respectively, whereas $K_{23}$
parameterizes the $SU(1,1)/U(1)\times SU(2)/U(1)$ \Km\ -- see
\cref{su11}.

\renewcommand{\arraystretch}{1.5}
\begin{table}
\begin{center}
\begin{tabular}{|c|c|c|c|c|}\hline
$i$ &$\hi(X)$ & $\hi''(0)$&\multicolumn{2}{c|}{$\widehat m_
s^2/\Hhi^2$} \\\cline{4-5}
&&& {$K=\kca$}&{$K=\kba$} \\\hline\hline
$1$&$X$&$0$&$-2+2^n/\ft^2$&$3\cdot2^{n-1}/\ft^2$\\\hline
$2$&$e^X-1$&$1$&$2^\frac{4-n}{2}\ck\sg^n-2+{2^{n}}/{\ft^2}$&$-6+3\cdot2^{n-1}/\ft^2$\\
$3$&$\ln(X+1)$&$-1$&$-2(1+2^{1-n/2}\ck\sg^n)$&$6(1+2^{n-1}/\ft^2)$\\\hline
$4$&$-\cos(\arcsin1+X)$&$0$&$-2(1-2^{n-1}/\ft^2)$&$3\cdot2^{n-1}/\ft^2$\\
$5$&$\sin(\arccos1+X)$&$0$&$-2(1-2^{n-1}/\ft^2)$&$3\cdot2^{n-1}/\ft^2$\\
$6$&$\tan\lf X\rg$&$0$&$-2(1-2^{n-1}/\ft^2)$&$3\cdot2^{n-1}/\ft^2$\\
$7$&$-\cot(\arcsin1+X)$&$0$&$-2(1-2^{n-1}/\ft^2)$&$3\cdot2^{n-1}/\ft^2$\\\hline
$8$&$\cosh(\arcsin1+X)-\sqrt{2}$&$\sqrt{2}$&$2^\frac{5-n}{2}\ck\sg^n-2+{2^{n}}/{\ft^2}$&$3\cdot2^{n-1}/\ft^2-6\sqrt{2}$\\
$9$&$\sinh(X)$&$0$&$-2(1-2^{n-1}/\ft^2)$&$3\cdot2^{n-1}/\ft^2$\\
$10$&$\tanh(X)$&$0$&$-2(1-2^{n-1}/\ft^2)$&$3\cdot2^{n-1}/\ft^2$\\
$11$&$-\coth(\arcsinh1+X)+\sqrt{2}$&$-2\sqrt{2}$&${2^{n}}/{\ft^2}-2^\frac{7-n}{2}\ck\sg^n-2$&$3\cdot2^{n-1}/\ft^2+12\sqrt{2}$\\
\hline
\end{tabular}\end{center}
\caption{\sl\ftn  Definition of the various $\hi(X)$'s,
$\hi''(0)=d^2\hi(0)/dX^2$ and masses squared of the fluctuations of
$s$ and $\bar s$ along the inflationary trajectory in \Eref{inftr}
for $K=\kca$ and $\kba$.} \label{tab1}
\end{table}

\renewcommand{\arraystretch}{1}

To show that the proposed $K$'s are suitable for IGI, we
have to verify that they reproduce $\Vhi$ in \sEref{VJig}{b} when
$n=2$, and they ensure the stability of $S$ at zero. These
requirements are checked in the following two sections.

\subsection{Derivation of the Inflationary Potential} \label{sec:ks2}

Substituting $W$ of \Eref{Wn} and a choice for $K$ in \Eref{k32}
(with the $\hi$'s given in \Tref{tab1}) into \Eref{Vsugra}, we
obtain a potential suitable for IGI. The inflationary trajectory
is defined by the constraints
\beq \label{inftr} S=T-T^*=0,\>\>\mbox{or}\>\>\>s=\bar s=\th=0\eeq
where we have expanded $T$ and $S$ in real and imaginary parts as follows
\beq T= \frac{\phi}{\sqrt{2}}\,e^{i\th}\>\>\>\mbox{and}\>\>\>S=
\frac{s +i\bar s}{\sqrt{2}}\cdot\label{cannor} \eeq

Along the path of \Eref{inftr}, $\Ve$  reads
\beq \label{1Vhio}\Vhi=\Ve(\th=s=\bar s=0)=e^{K}K^{SS^*}\,
|W_{,S}|^2\,.\eeq
From \Eref{Wn} we get $W_{,S}=f-1/2$. Also, \Eref{k32} yields
\beq \label{eK}  e^K= \begin{cases} \big(2f+\hi(0)\big)^{-\nc}&\mbox{for}~~~K=\kca\\
e^{\hi(0)}/(2f)^{\nb} &\mbox{for}~~~K=\kba\,,\end{cases}\eeq
where we take into account that $f(T)=f^*(T^*)$ along the path of
\Eref{inftr}. Moreover, $K^{SS^*}=1/K_{SS^*}$ can be obtained from the \Kaa\
metric, which is given by
\beq \label{Kab1}\lf K_{\al\bbet}\rg=\diag\lf
K_{TT^*},K_{SS^*}\rg=
\begin{cases}
\diag\lf {\nc n^2/2\sg^2},\hi'(0)/\big(2f+\hi(0)\big)\rg&\mbox{for}~~~K=\kca\\
\diag\lf {\nb\,
n^2/2\sg^2},\hi'(0)\rg&\mbox{for}~~~K=\kba\,,\end{cases} \eeq
where a prime denotes a derivative w.r.t. $X$. Note that
$K_{TT^*}$ for $K=\kba$ (and $S\neq0$) does not involve the field
$S$ in its denominator, and so no geometric destabilization
\cite{renaux} can be activated, contrary to the $K=\kca$ case.
Inserting $W_{,S}$, and the results of \eqs{eK}{Kab1} into
\Eref{Vsugra}, we obtain
\beq \label{Vhi1} \Vhi=\frac{\ld^2(1-2f)^2}{\ck^2} \cdot
\begin{cases} \big(2f+\hi(0)\big)^{1-\nc}/\hi'(0)&\mbox{for}~~K=\kca
\\e^{\hi(0)}/(2f)^{\nb}\hi'(0)&\mbox{for}~~K=\kba\,.\end{cases}\eeq

Recall that $f\sim\sg^n$ -- see \Eref{K0}. Then $\Vhi$ develops a
plateau, with almost constant potential energy density, for $\ck\gg1$ and $\sg<1$ (or $\ck=1$
and $\sg\gg1$), if we impose the following conditions:
\beq 2n=\left.\begin{cases}n(\nc-1)&\mbox{for}~~~K=\kca\\
n\nb&\mbox{for}~~~K=\kba\end{cases}\right\}\>\Rightarrow\>
\begin{cases} \nc=3&\mbox{for}~~~K= \kca\\
\nb=2&\mbox{for}~~~K=\kba\,.\end{cases} \label{ncond}\eeq
This empirical criterion is very precise since the data on $\ns$
allows only tiny (of order $0.001$) deviations \cite{tamvakis}.
Actually, the requirement $\ck\gg1$ and the synergy between the
exponents in $W$ and $K$'s assist us to tame the well-known $\eta$
problem within SUGRA with a mild tuning. If we insert \Eref{ncond}
into \Eref{Vhi1} and compare the result for $n=2$ with
\sEref{VJig}{b} (replacing also $\ld^2$ with $\ld$), we see that
the two expressions coincide, if we set
\beq \hi(0)=0~~~\mbox{and}~~~\hi'(0)=1\,.\label{cond}\eeq
As we can easily verify the selected $\hi$ in \Tref{tab1} satisfy
these conditions. Consequently, $\Vhi$ in \Eref{Vhi1} and the
corresponding Hubble parameter $\He_{\rm I}$ take their final
form:
\beq \label{Vhi} \mbox{\sf\ftn (a)}\>\>\>\Vhi=\frac{\ld^2\fp^2}{4
\ck^4 \sg^{2n}}\>\>\>\mbox{and}\>\>\> \mbox{\sf\ftn
(b)}\>\>\>\He_{\rm
I}={\Vhi^{1/2}\over\sqrt{3}}={\ld\fp\over2\sqrt{3}\ck^2\phi^n}\,,\eeq
with $\fp=2^{n/2-1} -\ck\sg^n<0$ reducing to that defined in
\sEref{VJig}{b}. Based on these expressions, we investigate in
\Sref{sec:inf} the dynamics and predictions of IGI.

\subsection{Stability of The Inflationary Trajectory} \label{stab}

\renewcommand{\arraystretch}{1.5}
\begin{table}
\begin{center}
\begin{tabular}{|c|c|c|c|c|}\hline
{\sc Fields} &{\sc Eigen-} & \multicolumn{3}{c|}{\sc Masses
Squared}\\\cline{3-5}
&{\sc states}&& {$K=\kca$}&{$K=\kba$} \\\hline\hline
1 real scalar & $\widehat\theta$& $\widehat m_{\theta}^2/\Hhi^2$&
$4(2^{n-2}-\ck\sg^n\fp)/\fp^2$
&$6(2^{n-2}-\ck\sg^n\fp)/\fp^2$\\\cline{2-5}
1 complex& $\widehat s, \widehat{\bar{s}}$ & $\widehat m_
s^2/\Hhi^2$&$2^{n}/\fp^2-2$&$3\cdot2^{n-1}/\fp^2$\\
%
scalar & & &$+2^{2-n/2}\ck\sg^n\hi''(0)$&$-6\hi''(0)$\\\hline
$2$ Weyl spinors & $\what \psi_\pm $ & $\what
m^2_{\psi\pm}/\Hhi^2$ &$2^n/\ft^{2}$ &$6\cdot2^{n-3}/\ft^{2}$
\\\hline
\end{tabular}
\end{center}
\caption{\sl\ftn  Mass-squared spectrum for $K=\kca$ and $\kba$
along the inflationary trajectory in \Eref{inftr}.} \label{tab2}
\end{table}
\renewcommand{\arraystretch}{1.}

We proceed to check the stability of the direction in \Eref{inftr} w.r.t. the
fluctuations of the various fields. To this end, we have to
examine the validity of the extremum and minimum conditions, i.e.,
\beq \mbox{\sf\ftn (a)}\>\>\>\left.{\partial
\Vhi\over\partial\what{\bar{z}}^\al}\right|_{s=\bar s=\th=0}=0\>\>\>
\mbox{and}\>\>\>\mbox{\sf\ftn (b)}\>\>\>\what m^2_{\bar
z^\al}>0\>\>\>\mbox{with}\>\>\>\bar z^\al=\th,s,\bar
s.\label{Vcon} \eeq
Here $\what m^2_{\bar z^\al}$ are the eigenvalues of the mass
matrix with elements
\beq \label{wM2}\what
M^2_{\al\bt}=\left.{\partial^2\Vhi\over\partial\what{\bar{z}}^\al\partial\what{\bar{z}}^\beta}\right|_{s=\bar s=\th=0}
\mbox{with}\>\>\>\bar z^\al=\th,s,\bar s\eeq
and a hat denotes the EF canonically normalized field. The canonically normalized fields can be
determined if bring the kinetic terms of the various scalars
in \Eref{Saction1} into the following form
\beqs\beq \label{K3} K_{\al\bbet}\dot z^\al \dot
z^{*\bbet}=\frac12\lf\dot{\se}^{2}+\dot{\what
\th}^{2}\rg+\frac12\lf\dot{\what s}^2 +\dot{\what{\overline
s}}^2\rg,\eeq
where a dot denotes a derivative w.r.t. the JF cosmic time. Then the
hatted fields are defined as follows
\beq  \label{cannor3b} {d\widehat \sg\over
d\sg}=J=\sqrt{K_{TT^*}},\>\>\> \what{\th}=
J\,\th/\phi,\>\>\>\mbox{and}\>\>\>(\what s,\what{\bar
s})=\sqrt{K_{SS^*}} {(s,\bar s)}\,,\eeq
where by virtue of
\eqs{ncond}{cond}, the \Kaa\ metric of \Eref{Kab1} reads
\beq\label{Kab}\lf K_{\al\bbet}\rg=\diag\lf{K_{TT^*}},
{K_{SS^*}}\rg=\begin{cases}
\lf{3n^2/2\sg^2},2^{n/2-1}/\ck\sg^n\rg&\mbox{for}~~~K=\kca\\
\lf{n^2/\sg^2},1\rg&\mbox{for}~~~K=\kba\,.\end{cases}\eeq \eeqs

Note that the spinor components $\psi_T$ and $\psi_S$
of the superfields $T$ and $S$ are normalized
in a similar way, i.e., $\what\psi_{T}=\sqrt{K_{TT*}}\psi_{\Phi}$
and $\what\psi_{S}=\sqrt{K_{SS^*}}\psi_{S}$. In practice,
we have to make sure that all the $\what m^2_{\bar z^\al}$'s are
not only positive, but also greater than $\Hhi^2$ during the last
$50-60$ e-foldings of IGI. This guarantees that the observed
curvature perturbation is generated solely by $\sg$, as assumed in
\Eref{Prob}. Nonetheless, two-field inflationary models which
interpolate between the \str\ and the quadratic model have been
analyzed in \cref{2field}. Due to the large effective masses that
the scalars acquire during IGI, they enter a phase of damped
oscillations about zero. As a consequence, the $\sg$ dependence in
their normalization -- see \Eref{cannor3b} -- does not affect
their dynamics.

We can readily verify that \sEref{Vcon}{a} is satisfied for all the three
$\bar z^\al$'s. Regarding \sEref{Vcon}{b}, we diagonalize $\what
M^2_{\al\bt}$, \Eref{wM2}, and we obtain the scalar mass
spectrum along the trajectory of \Eref{inftr}. Our
results are listed in \Tref{tab1} together with the masses squared
$\what m^2_{\psi\pm}$ of the chiral fermion eigenstates
$\what \psi_\pm=(\what{\psi}_{T}\pm \what{\psi}_{S})/\sqrt{2}$.
From these results, we deduce the following:

\begin{itemize}

\item For both classes of $K$'s in \Eref{k32}, \sEref{Vcon}{b} is satisfied
for the fluctuations of $\what\theta$, i.e. $\what m_{\th}^2>0$,
since $\ft<0$. Moreover, $\what m_{\th}^2\gg\Hhi^2$ because
$\ck\gg1$.

\item When $K=\kca$ and $\hi''(0)=0$, we obtain $\what m_{s}^2<0$.
This occurs for $i=1,4,...,7,9$ and $10$, as shown in \Tref{tab1}.
For $i=1$, our result reproduces those of similar models
\crefs{linde1, nIG, nmH, nmHkin}.  The stability problem can be
cured if we include in $\kca$ a higher order term of the form
$\kx|S|^4$ with $\kx\sim1$, or assuming that $S^2=0$ \cite{nil}.
However, a probably simpler solution arises if we take into
account the results accumulated in \Tref{tab2}. It is clear that the condition
$\what m_{s}^2>\Hhi^2$ can be satisfied when $\hi''(0)>0$ with
$|\hi''(0)|\geq1$. From \Tref{tab1}, we see that this is
the case for $i=2$ and $8$.

\item When $K=\kba$ and $\hi''(0)=0$, we obtain $\what m_{s}^2>0$,
but $\what m^2_{s}<\Hhi^2$. Therefore, $S$ may seed
inflationary perturbations, leading possibly to large
non-gaussianities in the CMB, contrary to observations. From the
results listed in \Tref{tab2}, we see that the condition $\what
m^2_{s}\gg\Hhi^2$ requires $\hi''(0)<0$ with $|\hi''(0)|\geq1$.
This occurs for $i=3$ and $11$. The former case was examined in \cref{su11}.

\end{itemize}

To highlight further the stabilization of $S$ during and after IGI
we present in \Fref{fig} $\what m^2_{s}/\Hhi^2$
as a function of $\sg$ for the various acceptable $K$'s
identified above. In particular, we fix $n=2$ and $\sgx=1$,
setting $K=K_{32}$ or $K=K_{38}$ in \sFref{fig}{a} and
$K=K_{23}$ or $K=K_{2,11}$ in \sFref{fig}{b}. The
parameters of the models ($\ld$ and $\ck$) corresponding to these choices are
listed in third and fifth rows of \Tref{tab3}. Evidently
$\what m^2_{s}/\Hhi^2$ remain larger than unity for
$\sgf\leq\sg\leq\sgx$, where $\sgx$ and $\sgf$ are also depicted.
However, in \sFref{fig}{b} $\what m^2_{s}/\Hhi^2$ exhibits a
constant behavior and increases sharply as $\sg$ decreases
below $0.2$. On the contrary, $\what m^2_{s}/\Hhi^2$ in
\sFref{fig}{a} is an increasing function of $\sg$ for
$\sg\gtrsim0.2$, with a clear minimum at $\sg\simeq0.2$.  For
$\sg\lesssim0.2$, $\what m^2_{s}/\Hhi^2$ increases drastically as
in \sFref{fig}{b} too.

\begin{figure}[!t]\vspace*{-.12in}
\hspace*{-.19in}
\begin{minipage}{8in}
\epsfig{file=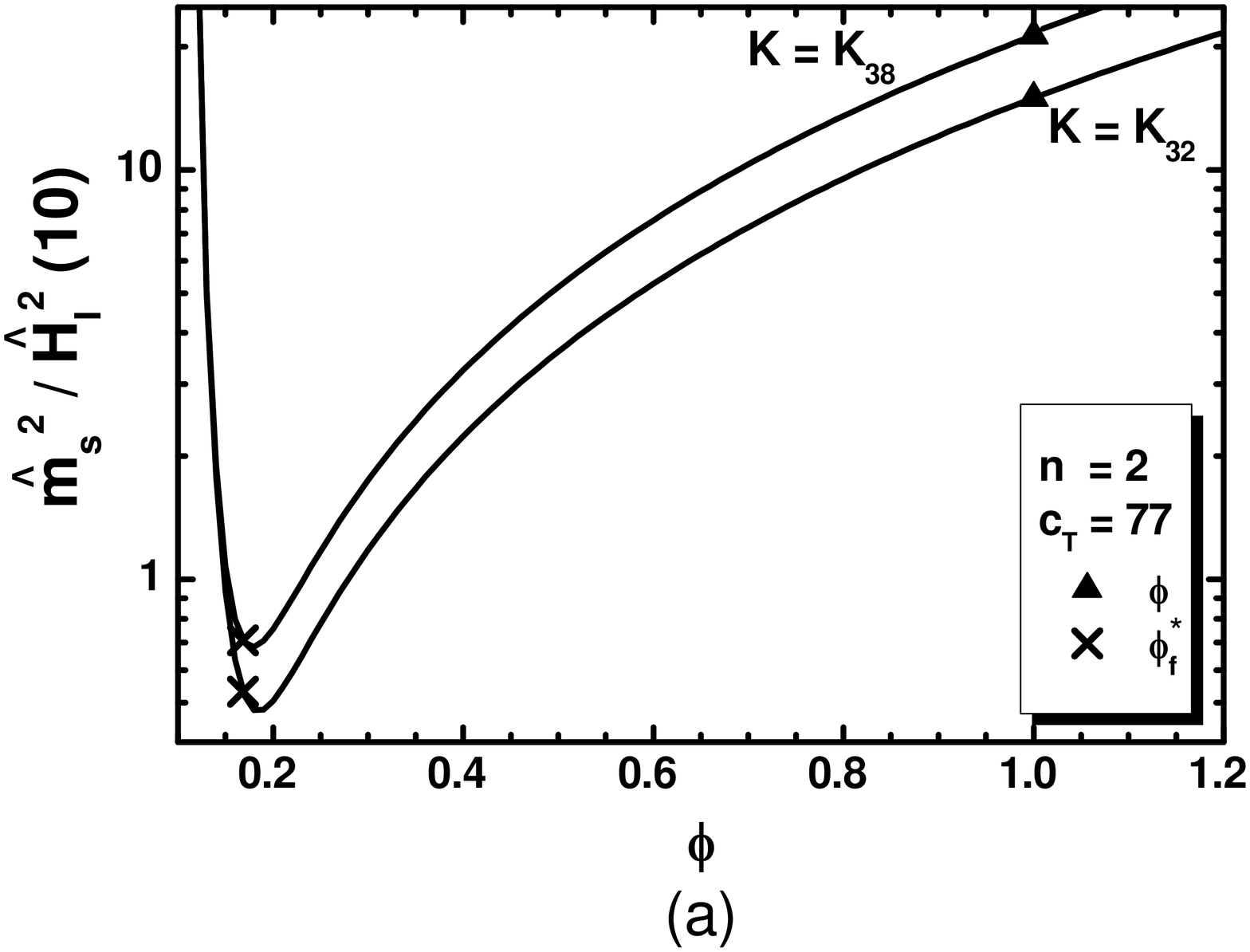,height=3.6in,angle=-90}
\hspace*{-1.2cm}
\epsfig{file=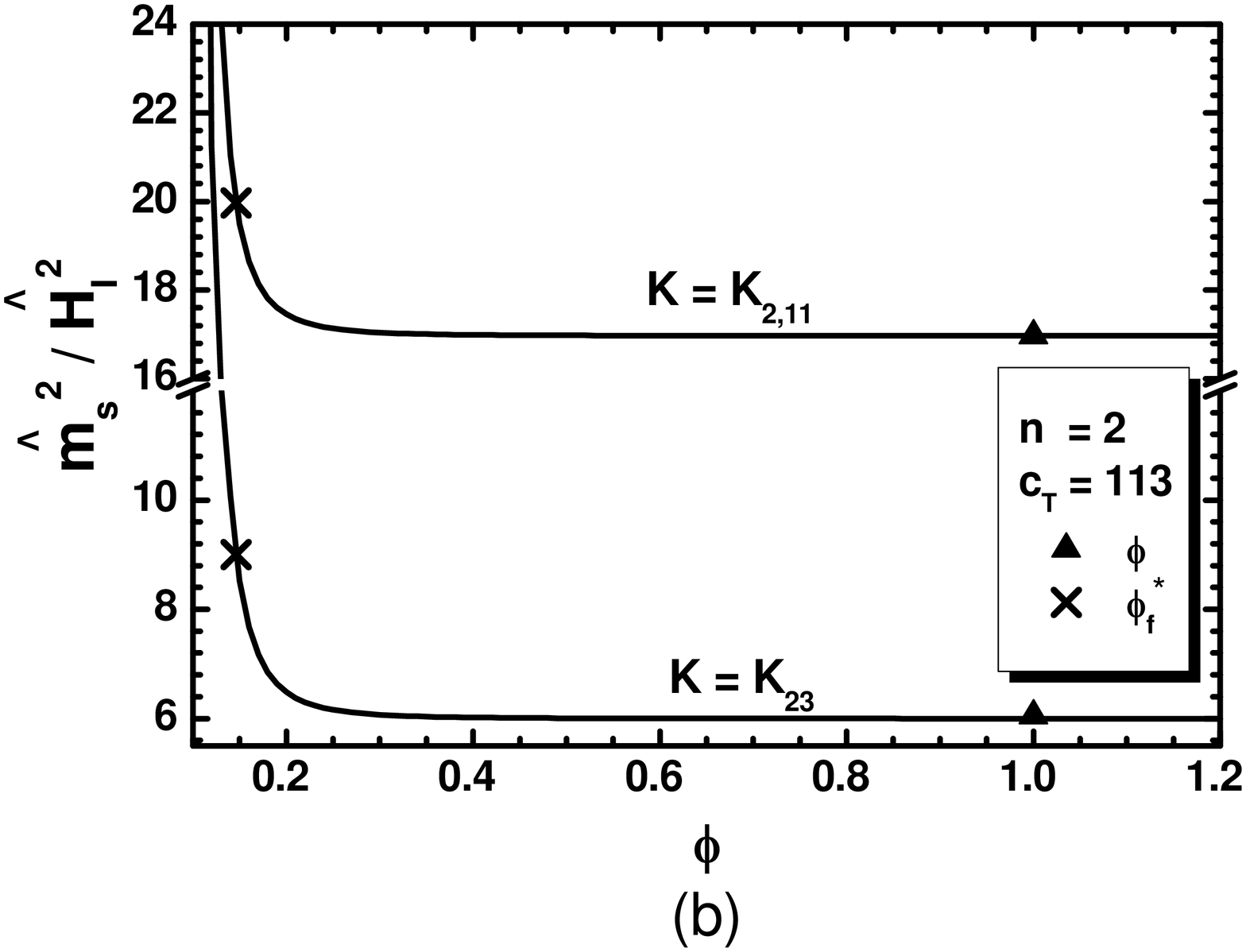,height=3.6in,angle=-90} \hfill
\end{minipage}
\hfill \caption[]{\sl \small The ratio $\what m^2_{s}/\Hhi^2$ as a
function of $\sg$  for $n=2$ and $\sgx=1$. We set {\sffamily\ssz
(a)} $K=K_{32}$ or $K=K_{38}$ and {\sffamily\ssz (b)} $K=K_{23}$
or $K=K_{2,11}$. The values corresponding to $\sgx$ and $\sgf$ are
also depicted.}\label{fig}
\end{figure}

Employing the well-known Coleman-Weinberg formula \cite{cw}, we
find from the derived mass spectrum -- see \Tref{tab1} --  the
one-loop radiative corrections, $\dV$, to $\Vhi$, depending on
renormalization group mass scale $\Lambda$. It can be verified
that our results are insensitive to $\dV$, provided that
$\Lambda$ is determined by requiring $\dV(\sgx)=0$ or
$\dV(\sgf)=0$. A possible dependence of the results on the choice
of $\Lambda$ is totally avoided \cite{nIG} thanks to the smallness
of $\dV$, for $\Lambda\simeq(1-1.8)\cdot10^{-5}$ -- see
\Sref{sec:res} too. These conclusions hold even for $\sg>1$.
Therefore, our results can be accurately reproduced by using
exclusively $\Vhi$ in \sEref{Vhi}{a}.

\section{Analysis of SUGRA Inflation}\label{sec:inf}

Keeping in mind that for $K=\kca$ [$K=\kba$] the values $i=2$ and
$8$ [$i=3$ and $11$] lead to the stabilization of $S$ during
and after IGI, we proceed with the computation of the
inflationary observables for the SUGRA models considered above.
Since the precise choice of the index $i$ does not influence our
outputs, here we do not specify henceforth the allowed $i$ values.
We first present, in \Sref{sec:inf1}, analytic results which are
in good agreement with our numerical results displayed in \Sref{sec:res}.
Finally we investigate the UV behavior of the models in
\Sref{sec:uv}.

\subsection{Analytical Results} \label{sec:inf1}

The duration of the IGI is controlled by the slow-roll parameters,
which are calculated to be
\beq \label{sr1}\lf\eph,\what\eta\rg=
\begin{cases}
\lf{2^n}/{3 \ft^2},{2^{1+n/2}(2^{n/2}-\ck\sg^n)}/{3
\ft^2}\rg&\mbox{for}~~~K=\kca\\
\lf{2^{n-2}}/{\ft^2},{2^{n/2}(2^{n/2}-\ck\sg^n)}/{\ft^2}\rg&\mbox{for}~~~K=\kba\,.\end{cases}
\eeq
The end of inflation is triggered by the violation of the $\what
\epsilon$ condition when $\sg=\sgf$ given by
\beqs\beq \what\epsilon\lf\sgf\rg=1\>\Rightarrow\>
\sgf\simeq\sqrt{2}\cdot\begin{cases}
\lf(1+2/\sqrt{3})/2\ck\rg^{1/n}&\mbox{for}~~~K=\kca\\
\lf(1+\sqrt{2})/2\ck\rg^{1/n}&\mbox{for}~~~K=\kba\,.\end{cases}
\label{sgap}\eeq
The violation of the $\what \eta$ condition occurs when
$\sg=\wtilde\sg_{f}<\sgf$:
\beq
\what\eta\lf\wtilde\sgf\rg=1\>\Rightarrow\>\wtilde\sgf\simeq\sqrt{2}\cdot\begin{cases}
\lf{5/6\ck}\rg^{1/n}&\mbox{for}~~~K=\kca\\
\lf{\sqrt{3}/2\ck}\rg^{1/n}&\mbox{for}~~~K=\kba\,.\end{cases}
\label{sgap1}\eeq\eeqs

Given $\sgf$, we can compute $\Ns$ via \Eref{Nhi}:
\beq \label{N*}
\Ns={\kp\over2}\lf2^{1-n/2}\ck\lf{\sgx^n-\sgf^n}\rg-n\ln\frac{\sgx}{\sgf}\rg
~~\mbox{with}~~~\kp=\begin{cases} 3/2&\mbox{for}~~~K=\kca\\
1&\mbox{for}~~~K=\kba\,.\end{cases} \eeq
Ignoring the logarithmic term and taking into account that
$\sgf\ll\sgx$, we obtain a relation between $\sgx$ and $\Ns$:
\beqs\beq\label{s*} \sgx\simeq\sqrt[n]{2^{n/2}\Ns/\kp\ck}.\eeq
Obviously, \FHI, consistent with \sEref{subP}{b}, can be achieved if
\beq \label{fsub} \sgx\leq1\>\>\>\Rightarrow\>\>\>\ck\geq
2^{n/2}\Ns/\kp\,.\eeq\eeqs
Therefore, we need relatively large $\ck$'s, which increase with
$n$. On the other hand, $\sex$ remains \trns, since solving the
first relation in \Eref{cannor3b} w.r.t. $\sg$ and inserting
\Eref{s*}, we find
\beq \sex\simeq\se_{\rm
c}+\sqrt{\kp}\ln(2\Ns/\kp)\simeq\begin{cases}
5.2&\mbox{for}~~~K=\kca\\
4.6&\mbox{for}~~~K=\kba\,,\end{cases} \label{se*}\eeq
where the integration constant $\se_{\rm c}=0$ and, as in the
previous cases, we set $\Ns\simeq53$. Despite this fact, our
construction remains stable under possible corrections from higher
order terms in $\fk$, since when these are expressed in terms of
initial field $T$, they can be seen to be harmless for $|T|\leq1$.

Upon substitution of \eqs{Vhi}{s*} into \Eref{Prob}, we find
\begin{equation}  \As^{1/2}\simeq\begin{cases}
{\ld(3-4\Ns)^2}/{96\sqrt{2}\pi\ck\Ns}&\mbox{for}~~~K=\kca\\
{\ld(1-2\Ns)^2}/{16\sqrt{3}\pi\ck\Ns}&\mbox{for}~~~K=\kba\,.\end{cases}\eeq
Enforcing \Eref{Prob}, we obtain a relation between $\ld$ and $\ck$,
which turns out to be independent of $n$. Indeed we have
\beq\ld\simeq \begin{cases}
6\pi\sqrt{2\As}\ck/\Ns\>\>\Rightarrow\>\>
\ck\simeq42969\ld\,&\mbox{for}~~~K=\kca\\
4\pi\sqrt{3\As}\ck/\Ns\>\>\Rightarrow\>\>
\ck\simeq52627\ld\,&\mbox{for}~~~K=\kba\,.\end{cases} \label{lan}
\eeq
Finally, substituting the value of $\sgx$ given in \Eref{s*} into \Eref{ns}, we estimate
the inflationary observables. For $K=\kca$ the results are given
in Eqs.~(\ref{nsig}) -- (\ref{rig}). For $K=\kba$ we obtain the
relations:
\beqs
\baq \label{ns2} && \ns=\frac{4\Ns(\Ns-3)-3}{(1-2\Ns)^2}\simeq1-{2\over\Ns}-{3\over\Ns^2}\simeq0.961;\\
\label{as2} && \as\simeq\frac{16\Ns(3+2\Ns)}{(2\Ns-1)^4}\simeq-{2\over\Ns^2}-{7\over\Ns^3}\simeq-0.00075;\\
\label{r2} &&
r\simeq\frac{32}{(1-2\Ns)^2}\simeq{8\over\Ns^2}+{8\over\Ns^3}\simeq0.0028\,.
\eaq\eeqs
These outputs are consistent with our results in \cref{su11} for
$m=n$ and $\na=\nb=2$ (in the notation of that reference).

\subsection{Numerical Results} \label{sec:res}

The analytical results presented above can be verified numerically.
The inflationary scenario depends on the following parameters --
see \eqs{Wn}{k32}:
$$n,\>\ck\>\>\>\mbox{and}\>\>\>\ld.$$
Note that the stabilization of $S$ with one of
$K_{32},K_{34},K_{23}$ and $K_{2,11}$ does not require any
additional parameter. Recall that we use $\Trh=4.1\cdot10^{-9}$
throughout and $\Ns$ is computed self-consistently for any $n$ via
\Eref{Nhi}. Our result is $\Ns\simeq53.2$. For given $n$, the
parameters above together with $\sgx$ can be determined by
imposing the observational constraints in Eqs.~(\ref{Nhi}) and
(\ref{Prob}). In our code we find $\sgx$ numerically, without the
simplifying assumptions used for deriving \Eref{s*}. Inserting it
into Eq.~(\ref{ns}), we extract the predictions of the models.

The variation of $\Vhi$ as a function of $\sg$ for two different
values of $n$ can be easily inferred from \Fref{Vn}. In
particular, we plot $\Vhi$ versus $\sg$ for $\sgx=1$, $n=2$ or
$n=6$, setting $K=\kca$ in \sFref{Vn}{a} and $K=\kba$ in
\sFref{Vn}{b}. Imposing $\sgx=1$ for $n=2$ amounts to
$(\ld,\ck)=(0.0017,77)$ for $K=\kca$ and $(\ld,\ck)=(0.0017,113)$
for $K=\kba$. Also, $\sgx=1$ for $n=6$ is obtained for
$(\ld,\ck)=(0.0068,310)$ for $K=\kca$ and $(\ld,\ck)=(0.0082,459)$
for $K=\kba$. In accordance with our findings in \Eref{fsub}, we
conclude that increasing $n$ {\sf\ftn (i)} requires larger $\ck$'s
and therefore, lower $\Vhi$'s to obtain $\sg\leq1$; {\sf\ftn (ii)}
larger $\sgf$ and $\vev{\phi}$ are obtained -- see \Sref{sec:uv}.
Combining \eqs{sgap}{lan} with \sEref{Vhi}{a}, we can conclude
that $\Vhi(\sgf)$ is independent of $\ck$ and to a considerable
degree of $n$.

\begin{figure}[!t]%
\hspace*{-.19in}
\begin{minipage}{8in}
\epsfig{file=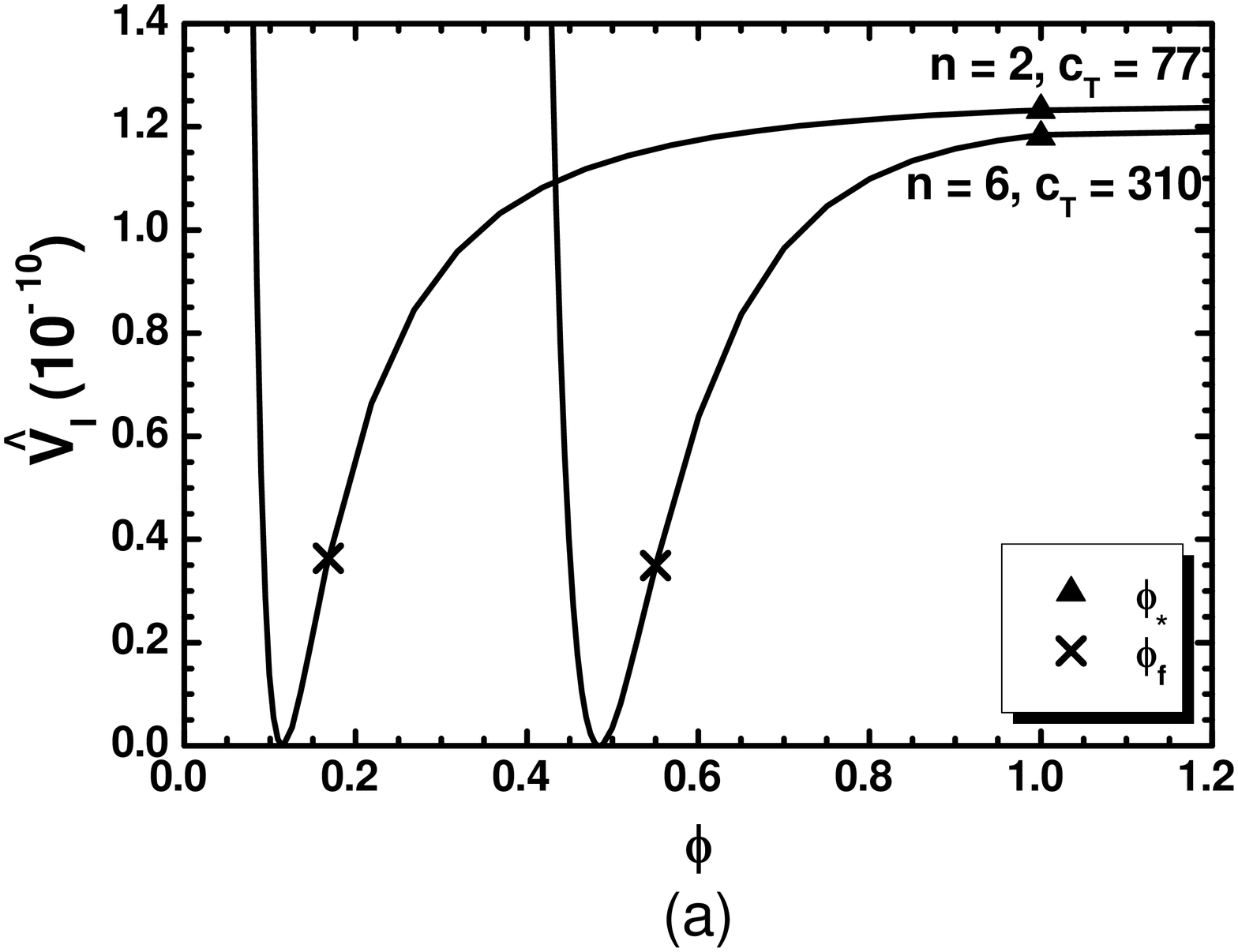,height=3.6in,angle=-90}
\hspace*{-1.2cm}
\epsfig{file=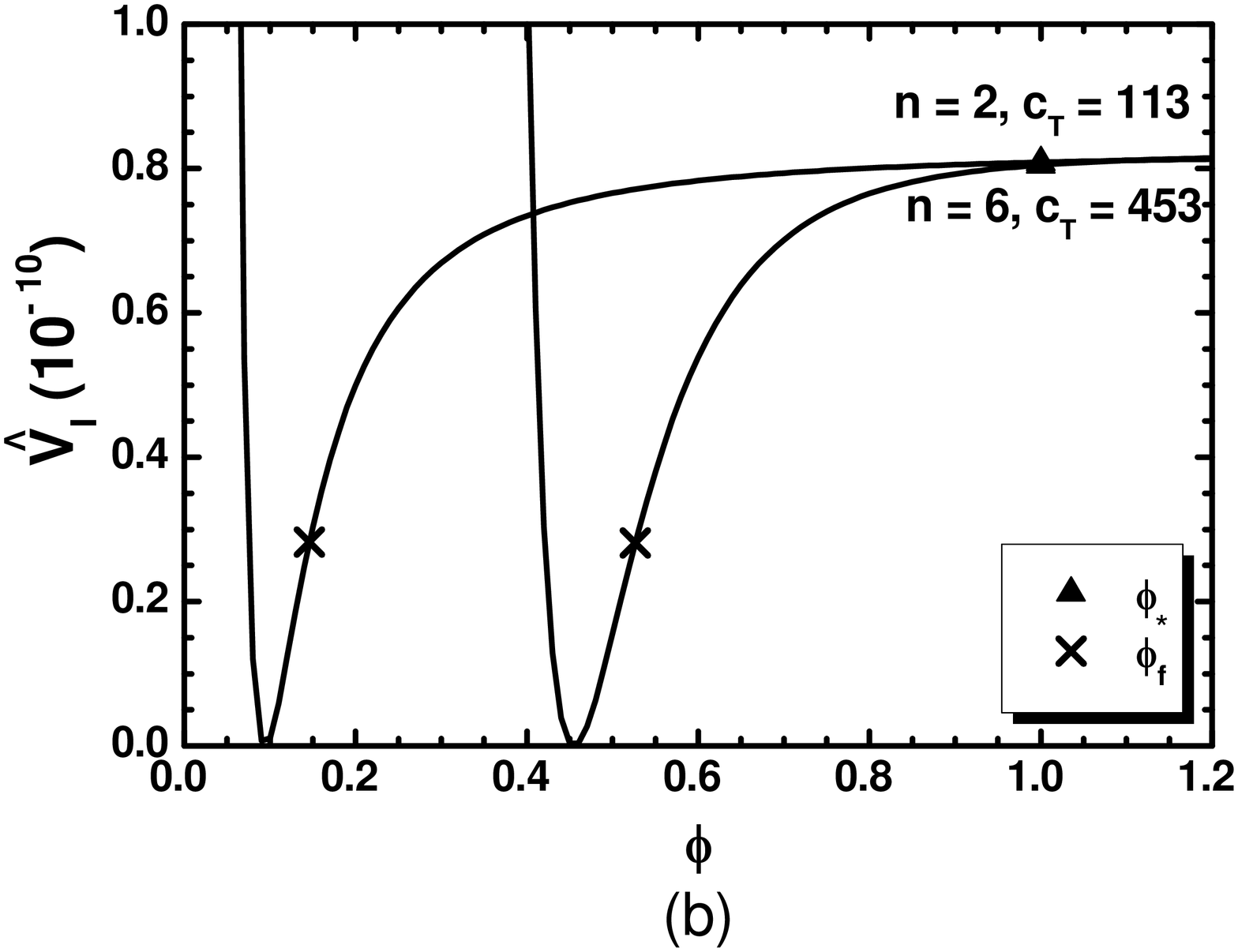,height=3.6in,angle=-90} \hfill
\end{minipage}
\hfill \caption[]{\sl The inflationary potential $\Vhi$ as a
function of $\sg$ for $\sgx=1$ and $n=2$ or $n=6$. We set
{\sffamily\ssz (a)} $K=\kca$ and {\sffamily\ssz (b)} $K=\kba$. The
values corresponding to $\sgx$ and $\sgf$ are also
depicted.}\label{Vn}
\end{figure}

Our numerical findings for $n=1,2$ and $3$ and $K=\kca$ or
$K=\kba$ are presented in \Tref{tab2}. In the two first rows, we
present results associated with Ceccoti-like models \cite{cec},
which are defined by $\ck=n=1$ and can not be made consistent
with the imposed $\mathbb{Z}_n$ symmetry or with
\Eref{subP}. We see that selecting $\sgx\gg1$, we attain solutions
that satisfy all the remaining constraints in \Sref{obs}.  For
the other cases, we choose a $\ck$ value so that $\sgx=1$.
Therefore, the presented $\ck$ is the minimal one, in agreement
with \Eref{fsub}.

In all cases shown in \Tref{tab2}, the model's predictions for
$\ns, \as$ and $r$ are independent of the input parameters. This
is due to the attractor behavior \cite{rena, gian,nIG} that
these models exhibit, provided that $\ck$ is large enough.
Moreover, these outputs are in good agreement with the analytical
findings of Eqs.~(\ref{nsig}) -- (\ref{rig}) for $K=\kca$ or
Eqs.~(\ref{ns2}) -- (\ref{r2}) for $K=\kba$. On the other hand,
the presented $\ck, \ld, \sgx$ and $\sgf$ values depend on $n$ for
every selected $K$. The resulting $\ns\simeq0.964$ is close to its
observationally central value; $r$ is of the order of $0.001$, and
$|\as|$ is negligible. Although the values of $r$ lie one order of
magnitude below the central value of the present combined \bcp\
and \plk\ results \cite{gws}, these are perfectly consistent with
the $95\%$ c.l. margin in \Eref{nswmap}. The values of $r$ for
$K=\kca$ or $K=\kba$ distinguish the two cases. The
difference is small, at the level of $10^{-3}$. However, it is
possibly reachable by the next-generation experiments. E.g., the
CMBPol experiment \cite{cmbpol} is expected to achieve a precision
for $r$ of the order of $10^{-3}$ or even $0.5\cdot10^{-3}$.
Finally, the renormalization scale $\Lambda$ of the
Coleman-Weinberg formula, found by imposing $\dV(\sgx)=0$, takes
the values $7.8\cdot10^{-5}$, $9.3\cdot10^{-5}$ $1.3\cdot10^{-5}$,
$2.1\cdot10^{-5}$ for $K_{32}$, $K_{38}$, $K_{23}$ and $K_{2,11}$
respectively. As a consequence, $\Lambda$ depends explicitly on
the specific choice of $i$ used for $\kca$ or $\kba$.

\renewcommand{\arraystretch}{1.25}
\begin{table}[!t]
\bec\begin{tabular}{|c||c|c|c||c|c||c|c|c|}\hline
{\sc K\"ahler} &  \multicolumn{8}{c|}{\sc Parameters}\\\cline{2-9}
{\sc Potential}&  \multicolumn{3}{c||}{\sc Input
}&\multicolumn{5}{|c|}{\sc Output}\\\cline{2-9}
$K$&$n$&$\ck$&$\sgx$&$\ld~(10^{-3})$&$\sgf$&$\ns$&$\as
(10^{-4})$&$r (10^{-3})$\\\hline\hline
$\kca$&$1$&$1$& $54.5$&$0.022$ &$1.5$&$0.964$ & $-6.3$& $3.6$\\
$\kba$&$1$&$1$& $80$&$0.028$ &$1.7$&$0.964$ & $-6.6$&
$2.5$\\\hline
$\kca$&$2$ &$77$& $1$&$1.7$&$0.17$& $0.964$ & $-6.7$& $3.7$\\
$\kca$&$3$ &$109$& $1$&$2.4$&$0.3$& $0.964$ & $-6.5$& $3.7$\\
$\kba$&$2$ &$113$& $1$&$2$&$0.15$& $0.964$ & $-6.7$& $2.5$\\
$\kba$&$3$ &$159$& $1$&$3$&$0.3$& $0.964$ & $-6.7$& $2.6$\\
\hline
\end{tabular}\eec
\hfill \caption[]{\sl\small  Input and output parameters of the
models which are compatible with \Eref{Nhi} for $\Ns=53.2$,
\Eref{Prob} and \Eref{nswmap}. } \label{tab3}
\end{table}

The overall allowed parameter space of the model for $n=2,3$ and
$6$ is correspondingly
\beqs\bea\label{res1} \fl &&~~~~~~~~~~ 77,105,310\lesssim
\ck\lesssim1.6\cdot10^5\>\>\>\mbox{and}\>\>\>(1.7,2.4,6.8)\cdot10^{-3}
\lesssim \ld\lesssim3.54\>\>\>\mbox{for}\>\>\>K=\kca\,; \\
\label{res2} \fl &&~~~~~~~~~~ 113,159,453\lesssim
\ck\lesssim1.93\cdot10^5\>\>\>\mbox{and}\>\>\>(2,2.9,8.2)\cdot10^{-3}
\lesssim \ld\lesssim3.54\>\>\>\mbox{for}\>\>\>K=\kba\,,\eea\eeqs
where the parameters are bounded from above as in \Eref{res4}.
Letting $\ld$ or $\ck$ vary within its allowed region above, we
obtain the values of $\ns, \as$ and $r$ listed in \Tref{tab3} for
$K=\kca$ and $\kba$ independently of $n$. Therefore, the inclusion
of the variant exponent $n>2$, compared to the non-SUSY model in
\Sref{igi}, does not affect the successful predictions of model.

\subsection{UV Behavior}\label{sec:uv}

Following the approach described in \Sref{obs}, we can verify that
the SUGRA realizations of IGI retain perturbative unitarity up to
$\mP$. To this end, we analyze the small-field behavior of the
theory, expanding ${\sf S}$ in \Eref{action1} about
\beq \vev{\sg}=2^{(n-2)/2n}\ck^{-1/n}\,, \label{vev}\eeq
which is confined in the ranges $(0.0026-0.1)$, $(0.021-0.24)$ and
$(0.17-0.48)$ for the margins of the parameters in
\eqs{res1}{res2}.

The expansion of ${\sf S}$ is performed in terms of $\dphi$ which
is found to be
\beq
\dphi=\vev{J}\dph\>\>\>\mbox{with}\>\>\>\vev{J}\simeq\sqrt{\kp}{n/\vev{\sg}}=
2^{(2-n)/2n}\sqrt{\kp}n\ck^{1/n}\,, \eeq
where $\kp$ is defined in \Eref{N*}. Note, in passing, that the
mass of $\dphi$ at the SUSY vacuum in \Eref{vevs} is given by
\beq\msn=\left\langle\Ve_{\rm I,\se\se}\right\rangle^{1/2}\simeq
\frac{\ld}{\sqrt{2\kp}\ck}
\simeq\frac{2\sqrt{6\As}\pi}{\Ns}\simeq1.25\cdot10^{-5}\,,\eeq
precisely equal to that found in \eqs{msstr}{msig}.  We observe
that $\msn$ is essentially independent of $n$ and $\kp$, thanks to
the relation between $\ld$ and $\ck$ in \Eref{lan}.

Expanding the second term in the r.h.s. of \Eref{Saction1} about
$\vev{\sg}$ with $J$ given by the first relation in
\Eref{cannor3b}, we obtain
\beqs\beq J^2 \dot\phi^2=
\lf1-\frac{2}{n\sqrt{\kp}}\what{\delta\phi}+\frac{3}{n^2\kp}\what{\delta\phi}^2-
\frac{4}{n^3}\kp^{-3/2}\dphi^3+\cdots\rg\dot\dphi^2\,.
\label{Jdph}\eeq
On the other hand, $\Vhi$ in \sEref{Vhi}{a} can be expanded about
$\vev{\phi}$ as follows
\beq\label{Vhiexp}
\Vhi\simeq\frac{\ld^2\se^2}{4\kp\ck^2}\lf1-\frac{n+1}{\sqrt{\kp}
n}\dphi+\lf1+n\rg\frac{11+7n}{12\kp
n^2}\dphi^2-\cdots\rg\,.\eeq\eeqs
Since the expansions above are $\ck$ independent, we infer that
$\Ld=1$ as in the other versions of Starobinsky-like inflation.
The expansions above for $K=\kca$ and $n=2$ reduce to those in
\eqs{Jig}{Veig}. Moreover, these are compatible with the ones
presented in \cref{nIG} for $K=\kca$ and those in \cref{su11} for
$K=\kba$ and $\na=2$. Our overall conclusion is that our models do
not face any problem with perturbative unitarity up to $\mP$.

\newpage

\section{Conclusions and Perspectives}\label{sec:con}

In this review we revisited the realization of Induced Gravity
Inflation (IGI) in both a non-supersymmetric and a Supergravity
(SUGRA) framework. In both cases the inflationary predictions
exhibit an attractor behavior towards those of Starobinsky model.
Namely, we obtained a spectral index $\ns\simeq(0.960-0.965)$ with
negligible running $\as$ and a tensor-to-scalar ratio
$0.001\lesssim r\lesssim0.005$. The mass of the inflaton turns out
be close to $3\cdot 10^{13}~\GeV$. It is gratifying that IGI can
be achieved for subplanckian values of the initial
(non-canonically normalized) inflaton, and the corresponding
effective theories are trustable up to Planck scale, although a
parameter has to take relatively high values. Moreover, the
one-loop radiative corrections can be kept under control.

In the SUGRA context this type of inflation can be incarnated
using two chiral supefields, $T$ and $S$, the superpotential in
\Eref{Wn}, which realizes easily the idea of induced gravity, and
several (semi)logarithmic K\"ahler potentials $\kca$ or $\kba$ --
see \Eref{k32}. The models are pretty much constrained upon
imposing two global symmetries -- a continuous R and a discrete
$\mathbb{Z}_n$ symmetry -- in conjunction with the requirement
that the original inflaton, $T$, takes \sub\ values. We paid
special attention to the issue of $S$ stabilization during IGI,
and worked out its dependence on the functional form of the
selected $K$'s with respect to $|S|^2$. More specifically, we
tested the functions $\hi(|S|^2)$, which appear in $\kca$ or
$\kba$ -- see \Tref{tab1}. We singled out $h_2(|S|^2)$ and
$h_{8}(|S|^2)$ for $K=\kca$ or $h_3(|S|^2)$ and $h_{11}(|S|^2)$
for $K=\kba$, which ensure that $S$ is heavy enough, and so well
stabilized during and after inflation. This analysis provides us
with new results that do not appear elsewhere in the literature.
Therefore, \str\ inflation realized within this SUGRA set-up
preserves its original predictive power, since no mixing between
$|T|^2$ and $|S|^2$ is needed for consistency in the considered
$K$'s -- cf. \cref{nIG,np1}.

It is worth emphasizing that the $S$-stabilization mechanisms
proposed in this paper can be also employed in other models of
ordinary \cite{linde1} or kinetically modified \cite{nMkin}
non-minimal chaotic (and Higgs) inflation driven by a gauge
singlet \cite{linde1,quad, nMkin} or non-singlet \cite{nmH,nmHkin}
inflaton, without causing any essential alteration to their
predictions. The necessary modifications involve replacing the
$|S|^2$ part of $K$ with $h_2(|S|^2)$, or $h_{8}(|S|^2)$ if we
have a purely logarithmic \Ka. Otherwise, the $|S|^2$ part can be
replaced by $h_3(|S|^2)$ or $h_{11}(|S|^2)$. Obviously, the last
case can be employed for logarithmic or polynomial $K$'s as
regards to the inflaton terms.

Let us, finally, remark that a complete inflationary scenario
should specify a transition to the radiation dominated era. This
transition could be facilitated in our setting \cite{R2r,
rehyoko,rehEllis} via the process of perturbative reheating,
according to which the inflaton after inflation experiences an
oscillatory phase about the vacuum, given by \Eref{ig2} for the
non-SUSY case or \Eref{vevs} for the SUGRA case. During this
phase, the inflaton can safely decay, provided that it couples to
light degrees of freedom in the lagrangian of the full theory.
This process is independent of the inflationary observables and
the stabilization mechanism of the non-inflaton field. It depends
only on the inflaton mass and the strength of the relevant
couplings. This scheme may also explain the origin of the observed
baryon asymmetry through non-thermal leptogenesis, consistently
with the data from the neutrino oscillations \cite{R2r}. It would
be nice to obtain a complete and predictable transition to the
radiation dominated era. An alternative graceful exit can be
achieved in the running vacuum models, as described in the fourth
paper of \cref{rvm}.

\section*{Disclaimer}

The authors declare that there is no conflict of interest
regarding the publication of this paper.


\def\prdn#1#2#3#4{{\sl Phys. Rev. D }{\bf #1}, no. #4, #3 (#2)}
\def\jcapn#1#2#3#4{{\sl J. Cosmol. Astropart.
Phys. }{\bf #1}, no. #4, #3 (#2)}

\section*{References}

\rhead[\fancyplain{}{ \bf \thepage}]{\fancyplain{}{\sl Starobinsky
Inflation: From non-SUSY to SUGRA Realizations}}
\lhead[\fancyplain{}{\sl References }]{\fancyplain{}{\bf
\thepage}} \cfoot{}


\begin{thebibliography}{99} {\ftn

\bibitem{guth} A.H. Guth, \prd{23}{1981}{347}; \\ A.D.~Linde, {\sl Phys.\ Lett.\ B }{\bf 108}, 389
(1982);\\  A.~Albrecht and P.J.~Steinhardt, {\sl Phys.\ Rev.\
Lett.} {\bf 48}, 1220 (1982).

\bibitem{R2} A.A.~Starobinsky, {\sl Phys.\ Lett.\ B }{\bf 91}, 99 (1980).

\bibitem{plin} P.A.R.~Ade {\it et al.}  [\plk\ Collaboration], \arxiv{1502.02114}.


\bibitem{gws}  P.A.R.~Ade {\it et al.} [{\sc BICEP2}/{\it Keck Array}
Collaborations], {\sl Phys.\ Rev.\ Lett.} {\bf 116}, 031302 (2016)
[\arxiv{1510.09217}].


\bibitem{nick1} K.S. Stelle, {\sl Phys. Rev. D } {\bf 16}, 953 (1977) ; K. S. Stelle,
{\sl Gen. Rel. Grav. }{\bf 9}, 353 (1978).


\bibitem{ketov}  S.V.~Ketov and A.A. Starobinsky, {\sl Phys. Rev. D }{\bf 83}, 063512
(2011) [\arxiv{1011.0240}]; \\S.V.~Ketov and N. Watanabe,
\jcap{03}{2011}{011} [\arxiv{1101.0450}]; \\S.V. Ketov and A.A.
Starobinsky, \jcap{08}{2012}{022} [\arxiv{1203.0805}]; \\S.V.
Ketov and S. Tsujikawa, \prd{86}{2012}{023529}
[\arxiv{1205.2918}].

\bibitem{ketov1} W.~Buchm\"uller, V.~Domcke and K.~Kamada, {\sl Phys.\ Lett.\ B
}{\bf 726}, 467 (2013) [\arxiv{1306.3471}];
\\F.~Farakos, A.~Kehagias and A.~Riotto, {\sl Nucl.\ Phys.\ }{\bf
B876} (2013) 187 [\arxiv{1307.1137}]; \\J.~Alexandre, N.~Houston
and N.E.~Mavromatos, {\sl Phys.\ Rev.\ D} {\bf 89}, 027703 (2014)
[\arxiv{1312.5197}]; \\K.~Kamada and J.~Yokoyama, {\sl Phys.\
Rev.\ D }{\bf 90}, 103520 (2014) [\arxiv{1405.6732}];\\
R.~Blumenhagen \etal, {\sl Phys.\ Lett.\ B} {\bf 746}, 217 (2015)
[\arxiv{1503.01607}]; \\ T.~Li, Z.~Li and D.V.~Nanopoulos,
\jhep{10}{2015}{138} [\arxiv{1507.04687}];\\ S.~Basilakos,
N.E.~Mavromatos and J.~Sola, {\sl Universe} {\bf 2}, no. 3, 14
(2016) [\arxiv{1505.04434}];

\bibitem{eno5} J.~Ellis, D.V.~Nanopoulos and K.A.~Olive,
{\sl Phys.\ Rev.\ Lett.\  }{\bf 111}, 111301 (2013); \\ {\sl
Erratum-ibid.\ } {\bf 111}, no. 12, 129902 (2013)
[\arxiv{1305.1247}].

\bibitem{eno7} J.~Ellis D.~Nanopoulos and K.~Olive, \jcap{10}{2013}{009} [\arxiv{1307.3537}].


\bibitem{linde} R.~Kallosh and A.~Linde, \jcap{06}{2013}{028} [\arxiv{1306.3214}].

\bibitem{zavalos} D. Roest, M. Scalisi and I. Zavala, \jcap{11}{2013}{007}
[\arxiv{1307.4343}].

\bibitem{matterlike} J.~Ellis, H.J.~He and Z.Z.~Xianyu,
{\sl Phys.\ Rev.\ D} {\bf 91}, no. 2, 021302 (2015)
[\arxiv{1411.5537}];\\ J.~Ellis \etal, \jcapn{11}{2016}{018}{11}
[\arxiv{1609.05849}]; \\ I.~Garg and S.~Mohanty, {\sl Phys.\
Lett.\ B} {\bf 751}, 7 (2015) [\arxiv{1504.07725}]. ,

\bibitem{ellis} J.~Ellis \etal, {\sl  Class.\ Quant.\ Grav.\  }{\bf 33}, no. 9, 094001 (2016)
[\arxiv{1507.02308}]; \\ G.~Chakravarty, G.~Lambiase and
S.~Mohanty, \arxiv{1607.06325}.



 \bibitem{tamvakis} A.B.~Lahanas and K.~Tamvakis, {\sl Phys.\
Rev.\ D }{\bf 91}, no. 8, 085001 (2015) [\arxiv{1501.06547}].

\bibitem{R2r} C.~Pallis, \jcap{04}{2014}{024}
[\arxiv{1312.3623}].


\bibitem{gian} G.F. Giudice and H.M. Lee, \plb{733}{2014}{58}
[\arxiv{1402.2129}].



\bibitem{nIG} C.~Pallis, \jcap{08}{2014}{057} [\arxiv{1403.5486}].


\bibitem{rena} R. Kallosh, {\sl Phys.\ Rev.\ D }{\bf 89}, no. 8, 087703 (2014)
[\arxiv{1402.3286}].



\bibitem{old} F.S. Accetta, D.J. Zoller, and M.S. Turner, {\sl Phys. Rev. D
}{\bf 31}, 3046 (1985); \\ D.S. Salopek, J. R. Bond and J.M.
Bardeen, {\sl Phys. Rev. D }{\bf 40}, 1753 (1989); \\ R. Fakir and
W.G. Unruh, {\sl Phys. Rev. D }{\bf 41}, 1792 (1990).

\bibitem{higgsflaton} J.L.~Cervantes-Cota and H.~Dehnen, \prd{51}{1995}{395} [\astroph{9412032}];\\
N. Kaloper, L. Sorbo and J. Yokoyama, {\sl Phys. Rev. D }{\bf 78},
043527 (2008)[\arxiv{0803.3809}]; \\ A. Cerioni, F. Finelli, A.
Tronconi and G. Venturi, \prd{81}{2010}{123505}
[\arxiv{1005.0935}].

\bibitem{jones} K.~Kannike \etal,
\jhep{05}{2015}{065} [\arxiv{1502.01334}];\\ M.B.~Einhorn and
D.R.T.~Jones, \jhep{01}{2016}{019} [\arxiv{1511.01481}].


\bibitem{zee} A. Zee, {\sl Phys. Rev. Lett. }{\bf 42}, 417
(1979); H.~Terazawa, {\sl Phys.\ Lett.\ B} {\bf 101}, 43 (1981).


\bibitem{sm1} J.L.~Cervantes-Cota and H.~Dehnen, \npb{442}{1995}{391}
[\astroph{9505069}];\\ F.L.~Bezrukov and M.~Shaposhnikov,
\plb{659}{2008}{703} [\arxiv{0710.3755}].


\bibitem{nmi} C.~Pallis, \plb{692}{2010}{287} [\arxiv{1002.4765}].



\bibitem{atroest} R. Kallosh, A. Linde and D. Roest,
{\sl Phys. Rev. Lett.} {\bf 112}, 011303 (2014)
[\arxiv{1310.3950}].

\bibitem{linde1}  M.B.~Einhorn and D.R.T.~Jones,
\jhep{03}{2010}{026} [\arxiv{0912.2718}]; \\ S.~Ferrara \etal,
\prd{83}{2011}{025008} [\arxiv{1008.2942}]; \\ C.~Pallis and
N.~Toumbas, \jcap{02}{2011}{019} [\arxiv{1101.0325}].

\bibitem{nmH}  M.~Arai, S.~Kawai and N.~Okada, {\sl Phys. Rev. D}
{\bf 84},1 23515 (2011) [\arxiv{1107.4767}]; \\ C.~Pallis and
N.~Toumbas, \jcap{12}{2011}{002} [\arxiv{1108.1771}]; \\ M.B.
Einhorn and D.R.T. Jones, \jcap{11}{2012}{049}
[\arxiv{1207.1710}].



\bibitem{quad} C. Pallis and Q. Shafi, \prd{86}{2012}{023523}
[\arxiv{1204.0252}];\\ C.~Pallis and Q.~Shafi, \jcap{03}{2015}{no.
03, 023} [\arxiv{1412.3757}].









\bibitem{cutoff} J.L.F.~Barbon and J.R.~Espinosa,
\prd{79}{2009}{081302} [\arxiv{0903.0355}]; \\ C.P.~Burgess,
H.M.~Lee, and M.~Trott, \jhep{07}{2010}{007} [\arxiv{1002.2730}].


\bibitem{riotto} A.~Kehagias, A.M.~Dizgah, and A.~Riotto, \prd{89}{2014}{043527}
[\arxiv{1312.1155}].



\bibitem{su11} C.~Pallis and N.~Toumbas, \jcap{05}{2016}{no. 05, 015}
[\arxiv{1512.05657}].

\bibitem{noscale} E.~Cremmer, S.~Ferrara, C.~Kounnas and D.V.~Nanopoulos,
{\sl Phys.\ Lett.\ B } {\bf 133}, 61 (1983); \\ J.R.~Ellis,
A.B.~Lahanas, D.V.~Nanopoulos and K.~Tamvakis, {\sl Phys.\ Lett.\
B } {\bf 134}, 429 (1984).

\bibitem{lahanas} A.B. Lahanas and D.V.~Nanopoulos, {\sl Phys.\ Rept. } {\bf 145}, 198 (1987).

\bibitem{rehyoko}  T.~Terada, Y.~Watanabe, Y.~Yamada and J.~Yokoyama,
\jhep{02}{2015}{105} [\arxiv{1411.6746}].

\bibitem{rehEllis} J.~Ellis, M.~Garcia, D.~Nanopoulos and
K.~Olive, \jcap{10}{2015}{003} [\arxiv{1503.08867}].


\bibitem{lee} H.M.~Lee, \jcap{08}{2010}{003} [\arxiv{1005.2735}].

\bibitem{np1} C.~Pallis, \jcap{10}{2014}{058} [\arxiv{1407.8522}]; \\
C.~Pallis, {\sl PoS CORFU} {\bf 2014}, 156 (2015)
[\arxiv{1506.03731}].


\bibitem{nMkin} C.~Pallis,~\prdn{91}{2015}{123508}{12} [\arxiv{1503.05887}];
\\ C.~Pallis, {\sl PoS PLANCK} {\bf 2015}, 095 (2015)
[\arxiv{1510.02306}]; \\ C.~Pallis, \arxiv{1611.07010}.


 \bibitem{nmHkin} G.~Lazarides and C.~Pallis, {\sl J. High Energy
Phys.} {\bf 11}, 114 (2015) [\arxiv{1508.06682}];\\ C.~Pallis,
{\sl Phys. Rev. D} {\bf 92}, no. 12, 121305(R) (2015)
[\arxiv{1511.01456}];\\ C.~Pallis, \jcapn{10}{2016}{037}{10}
[\arxiv{1606.09607}].




\bibitem{nick}  C.~Kounnas, D.~L\"ust and N.~Toumbas, {\sl Fortsch.\ Phys.}  {\bf 63}, 12 (2015)
[\arxiv{1409.7076}].




\bibitem{talk} C. Pallis, {\sl PoS CORFU} {\bf 2012}, 061 (2013)
[\arxiv{1307.7815}].


\bibitem{nil} I.~Antoniadis, E.~Dudas, S.~Ferrara and A.~Sagnotti,
{\sl Phys.\ Lett.\ B } {\bf 733}, 32 (2014) [\arxiv{1403.3269}].


\bibitem{rvm} J.A.S.~Lima, S.~Basilakos and J.~Sol\`a, {\sl Mon.\ Not.\ Roy.\ Astron.\ Soc.\  }{\bf 431}, 923 (2013)
[\arxiv{1209.2802}];\\   J.A.S.~Lima, S.~Basilakos and J.~Sol\` a,
{\sl Gen.\ Rel.\ Grav.\  }{\bf 47}, 40 (2015)
[\arxiv{1412.5196}];\\  J.~Sol\`a and A.~G\'omez-Valent, {\sl
Int.\ J.\ Mod.\ Phys.\ D }{\bf 24}, 1541003 (2015)
[\arxiv{1501.03832}]; \\ J.A.S.~Lima, S.~Basilakos and J.~Sol\`a,
{\sl Eur.\ Phys.\ J.\ C }{\bf 76}, no. 4, 228 (2016)
[\arxiv{1509.00163}]; \\  J.~Sol\`a, A.~G\'omez-Valent and J.~de
Cruz Pérez, {\sl Astrophys.\ J.\  }{\bf 836}, no. 1, 43 (2017)
[\arxiv{1602.02103}];\\   J.~Sol\`a, A.~G\'omez-Valent and J.~de
Cruz P\'erez, {\sl Astrophys.\ J.\  }{\bf 811}, L14 (2015)
[\arxiv{1506.05793}];\\  J.~Sol\`a, J.~de Cruz P\'erez,
A.~G\'omez-Valent and R.C.~Nunes, \arxiv{1606.00450}\\ J.~Sol\`a,
{\sl J.\ Phys.\ A }{\bf 41}, 164066 (2008) [\arxiv{0710.4151}].








\bibitem{review} D.H.~Lyth and
A.~Riotto, {\sl Phys.\ Rept.} {\bf 314}, 1 (1999) [{\tt\ftn
hep-ph/9807278}];  \\ G. Lazarides, {\sl J. Phys. Conf. Ser.} {\bf
53}, 528 (2006) [{\tt\ftn hep-ph/0607032}]; \\ A.~Mazumdar and
J.~Rocher, {\sl Phys.\ Rept. }{\bf 497}, 85 (2011)
[\arxiv{1001.0993}]; \\ J.~Martin, C.~Ringeval and V.~Vennin, {\sl
Physics of the Dark Universe} {\bf 5-6}, 75 (2014)
[\arxiv{1303.3787}].


\bibitem{turner} M.S.~Turner, {\sl Phys. Rev. D} {\bf 28}, 1243 (1983).

\bibitem{plcp} P.A.R. Ade \etal\ [Planck Collaboration], \arxiv{1502.01589}.



\bibitem{2field} J.~Ellis \etal, \jcap{01}{2015}{010} [\arxiv{1409.8197}]; \\ C.~van de Bruck and
L.E.~Paduraru, {\sl Phys.\ Rev.\ D} {\bf 92}, no. 8, 083513 (2015)
[\arxiv{1505.01727}]; \\ S.~Kaneda and S.V.~Ketov, {\sl Eur.\
Phys.\ J.\ C }{\bf 76}, no. 1, 26 (2016) [\arxiv{1510.03524}]; \\
G.~Chakravarty, S.~Das, G.~Lambiase and S.~Mohanty, {\sl Phys.\
Rev.\ D }{\bf 94}, no. 2, 023521 (2016) [\arxiv{1511.03121}].


\bibitem{renaux} S.~Renaux-Petel and K.~Turzynski, {\sl Phys.\ Rev.\ Lett.\ }{\bf 117}, no. 14, 141301 (2016)
[\arxiv{1510.01281}].








\bibitem{cw} S.R.~Coleman and E.J.~Weinberg, \prd{7}{1973}{1888}.

\bibitem{cec} S.~Cecotti, {\it Phys.\ Lett.\ B }{\bf 190}, 86
(1987).



\bibitem{cmbpol}  D. Baumann \etal, {\sl AIP Conf.Proc. }{\bf 1141}, 10
 (2009) [\arxiv{0811.3919}].

  }


\end{thebibliography}
\end{document}